\documentclass[prc,twocolumn,showpacs,preprintnumbers,amsmath,amssymb]{revtex4}

\usepackage{amsmath,amssymb,multirow,epsfig,bm,color}

\newcommand{\beq}{\begin{equation}}
\newcommand{\eeq}{\end{equation}}
\newcommand{\bea}{\begin{eqnarray}}
\newcommand{\eea}{\end{eqnarray}}

\newcommand{\benn}{\begin{displaymath}}
\newcommand{\eenn}{\end{displaymath}}
\newcommand{\angstr}{\rm\AA}

\begin{document}

\title{Spin effects in thermoelectric properties of Al and P doped zigzag silicene nanoribbons}

\author{K. Zberecki$^1$,  R. Swirkowicz$^1$, J. Barna\'s$^2$}
\affiliation{$^1$Faculty of Physics, Warsaw University of Technology, ul. Koszykowa 75, 00-662 Warsaw, Poland}
\affiliation{$^2$Faculty of Physics, Adam Mickiewicz University, ul. Umultowska 85, 61-614 Pozna\'n, Poland\\ and Institute of Molecular Physics, Polish Academy of Sciences, Smoluchowskiego 17, 60-179 Pozna\'n, Poland}
\date{\today}

\begin{abstract}
Electric and thermoelectric properties of silicene nanoribbons doped with Al and P impurity atoms are investigated theoretically for both antiparallel and parallel orientations of the edge magnetic moments. In the former case, appropriately arranged impurities can lead to a net magnetic moment and thus
also to spin thermoelectric effects. In the latter case, in turn, spin thermoelectric effects also occur in the absence of impurities. Numerical results based on {\it ab-initio} calculations show that the spin  thermopower can be considerably enhanced by the impurities.
\end{abstract}

\pacs{xxx}

\maketitle

\section{Introduction}

Since the discovery of graphene, one can observe increasing interest in other two-dimensional honeycomb structures. One of such materials is silicene -- a two-dimensional hexagonal lattice of silicon (Si) atoms. In contrast to graphene, silicene has
a buckled atomic structure, where the two triangular sublattices are slightly displaced in opposite directions normal to the atomic plane. Band structure calculations show that silicene is
a semimetal with zero energy gap and linear electronic spectrum  near the  $K$ points of the Brilouine zone~\cite{1}. Similarly to graphene, electrons
in the vicinity of the $K$ points (Dirac points) behave like massless fermions. Apart from strictly two-dimensional crystals of silicene,  also silicene nanoribbons have been fabricated recently~\cite{2,3,4,5}.

Theoretical investigations of silicene have revealed a number of its interesting properties, like for instance the spin Hall effect induced by spin-orbit interaction~\cite{6} or an electrically tunable energy gap~\cite{7,8}. The latter effect appears owing to the buckled atomic structure. Moreover, {\it ab-initio} numerical calculations  have shown that two electrically-controlled gapped Dirac
cones for nearly spin polarized states can exist when silicene is in a perpendicular electric field. Accordingly, an effective silicene-based spin
filter has been proposed, where spin polarization of electric current can be switched with external electric field ~\cite{9}. Since silicon plays a crucial role in the present-day electronics, integration of silicene into nanoelectronics seems to be more promising than that of graphene, and this possibility opens new perspectives for this
novel material ~\cite{10,11}. Therefore, detailed description and understanding of physical properties of silicene is currently of
great interest.

First-principle calculations based on the density functional theory (DFT) have shown that silicene surface is very reactive and can be
easily functionalized~\cite{12}. This functionalization, in turn, can significantly change basic properties of silicene. Structural,
electronic, magnetic and vibrational properties of silicene with adsorbed or substituted atoms of various types have been extensively studied in recent years~\cite{12,14}. The corresponding results show that the low-buckled lattice is stable in a wide range of doping~\cite{13}. Apart from this, both ad-atoms and substituted atoms induce characteristic modes in the phonon spectrum of silicene. A significant charge transfer between
ad-atoms and silicene has also been reported~\cite{12}.

Especially interesting are functionalization- and doping-induced modifications of transport properties. It has been shown that adsorption of alkali metals can transform silicene into a narrow gap semiconductor. On the other hand, by doping with transition-metal atoms, either semiconducting or metallic behavior can be obtained~\cite{14}.
Furthermore, the band gap in silicene can be tuned when it is functionalized with hydrogen~\cite{15,16}. As a result of semi-hydrogenation, i.e.
hydrogenation from one side only, ferromagnetic ordering can be induced, and the system exhibits then semiconducting properties with a direct
energy gap of the order of 1 eV ~\cite{17}. It is worth noting that itinerant magnetism mediated by holes has been also predicted for AlSi monolayers as well as for
AlSi armchair nanoribbons~\cite{18}.

The electronic structure can be also modified by externally induced strain. For instance, it has been predicted that compressive (tensile) strain can move the Dirac points in silicene below (above) the Fermi level, so the system can behave like n-type (p-type) doped one~\cite{19}.
Moreover, a sufficiently strong strain can reduce the thermal conductivity of silicene, as follows from non-equilibrium molecular dynamic simulations~\cite{20}.

Apart from two-dimensional silicene crystals, also nanoribbons of armchair (aSiNRs) or zigzag (zSiNRs) type have been widely studied in view of
potential applications in silicon-based electronics and spintronics devices. Similarly to graphene nanoribbons (GNRs), zSiNRs exhibit
edge magnetic ordering. {\it Ab-initio} calculations for  pristine zSiNRs show that antiferromagnetic (AFM) state, where magnetic moments of the two
edges of a nanoribbon are antiparallel, corresponds to the lowest energy~\cite{21,22}. Ferromagnetic (FM) ordering, in which magnetic
moments of the two edges are parallel, has a slightly higher energy and can be stable {\it eg} in an external magnetic field. In the
FM state, zSiNRs have metallic transport properties, whereas in the AFM configuration a wide gap opens and zSiNRs display semiconducting behavior. Electronic,
mechanical and magnetic properties of SiNRs have been studied recently by first principle methods~\cite{21,22,23,24,25}. In
particular, investigations of electron transport properties of zSiNRs have revealed  a giant magnetoresistance effect associated with transition from
the FM to AFM configuration ~\cite{24}. It has been also predicted that hydrogen-terminated zSiNRs in the presence of in-plane electrical
field can behave like a half-metallic ferromagnet with spin polarization up to 99$\%$ ~\cite{26}. When a local exchange field affects only one edge of zSINRs, an energy gap can be opened in one spin channel, whereas the second spin channel remains gapless
~\cite{27}. Spin gapless semiconductor behavior with 100$\%$ spin polarization has been also predicted for zSiNRs doped with B or N
atoms in edge positions ~\cite{28,29}. On the other hand, in aSiNRs with B/N substitutions at the edges,
semiconductor-metal transition
due to formation of a half-filled impurity band near the Fermi level has been  predicted~\cite{29}. Very recently, the interplay between
bulk and edge states induced by Rashba spin-orbit coupling in zSiNRs has been investigated in the presence of electrical field
~\cite{30}. It has been shown that the states with opposite velocities can open spin-dependent subgaps which remarkably influence spin polarized
transport properties.

Thermoelectric properties of nanoscopic systems are currently of great interest due to the possibility of heat to electrical energy
conversion at nanoscale, which is important for applications. Thermoelectric properties of aSiNRs and zSINRs have been investigated
by {\it ab-initio} methods based on DFT and non-equilibrium Green function formalism~\cite{21,22}. It has been shown that thermopower of
pristine zSiNRs in the low-energy AFM state can be considerably enhanced due to the presence of energy gaps. Moreover,
the thermopower strongly depends on the magnetic configuration and a considerable magnetothermopower  related to transition
from the AFM to FM state can be observed~\cite{22}. In ferromagnetic systems, interplay between the spin effects and thermoelectric properties can
lead to new spin related thermoelectric phenomena~\cite{31,32,33,34}. The most spectacular spin related thermoelectric effect is the spin
thermopower (spin Seebeck effect) which is a spin analog of the conventional thermopower. A considerable spin thermopower has been
predicted for pristine zSiNRs with FM ordering ~\cite{22}.

In the present paper we analyze electric and thermoelectric phenomena of zSiNRs doped with Al and P impurity atoms. It is
shown that the spin thrmopower can be considerably enhanced by impurities. In section 2 we present transmission function for zSiNRs. Thermoelectric properties in the AFM state are presented and discussed in section 3, while those in the FM state are considered in section 4. Summary and final conclusions are in section 5.

\section{Transmission in ${\rm z}$S${\rm i}$NR${\rm s}$ with A${\rm l}$ and P impurity atoms}

In this section we will present  numerical results on electronic transmission in zSiNRs with impurities, obtained
  by {\it ab-initio} numerical calculations
within the DFT Siesta code ~\cite{siesta1}. The nanoribbon edges were terminated with hydrogen
atoms to remove the dangling bonds, and pristine zSiNRs as well as those with  impurity atoms
of Al or P type, localized at different positions with respect to the nanoribbon edges, were considered.
The impurities were distributed periodically along the chain and localized (i) at one of the edges (PE configuration), (ii) at the nanoribbon center (PC configuration), and (iii)  in the middle between the edge and central
atoms of the nanoribbon (PM configuration). The elementary cell was adequately enlarged to include one impurity atom.
The spin-resolved energy-dependent transmission $T_\sigma(E)$ through nanoribbons was determined
within the non-equilibrium Green function (NEGF) method as implemented in the Transiesta code
~\cite{siesta2}. The structures were optimized until atomic forces converged to 0.02 eV/A.
The atomic double-polarized basis (DZP) was used and the grid mesh cutoff was set equal to
200 Ry. The generalized gradient approximation (GGA) with Perdrew-Burke-Ernzerhof parameterization
was applied for exchange-correlation part of the total energy functional ~\cite{pbe1}.
The performed calculations, similarly to those presented in Refs ~\cite{21,22}, show that
antiferromagnetic (AFM) state, where magnetic moments at one edge are antiparallel to those at the other edge,
is the most stable configuration (ground state) in pristine narrow zSiNRs.
Ferromagnetic (FM) state, in which the edge moments are all parallel, corresponds to
slightly higher energy. The energy difference between the  two magnetic states is equal to
0.02 eV for pristine nanoribbon containing $N$=6 zigzag chains. Thus, magnetic configuration of pristine zSiNRs can be
easily changed from the AFM state to the FM one, for instance  by an external magnetic field.

\subsection{Low energy state}

\begin{figure}[ht]
  \begin{center}
    \begin{tabular}{cc}
      \resizebox{42mm}{!}{\includegraphics[angle=0]{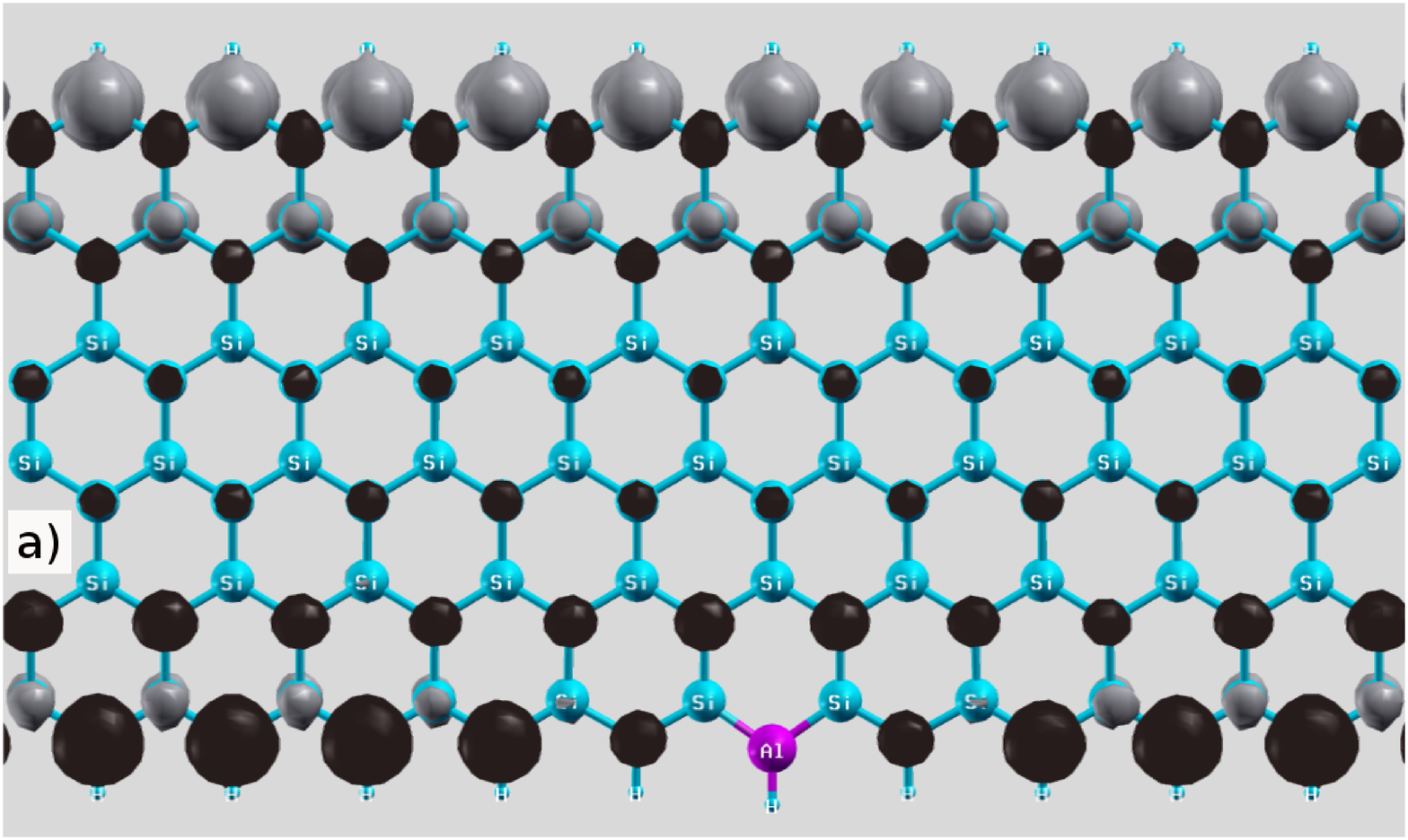}} &
      \resizebox{42mm}{!}{\includegraphics[angle=0]{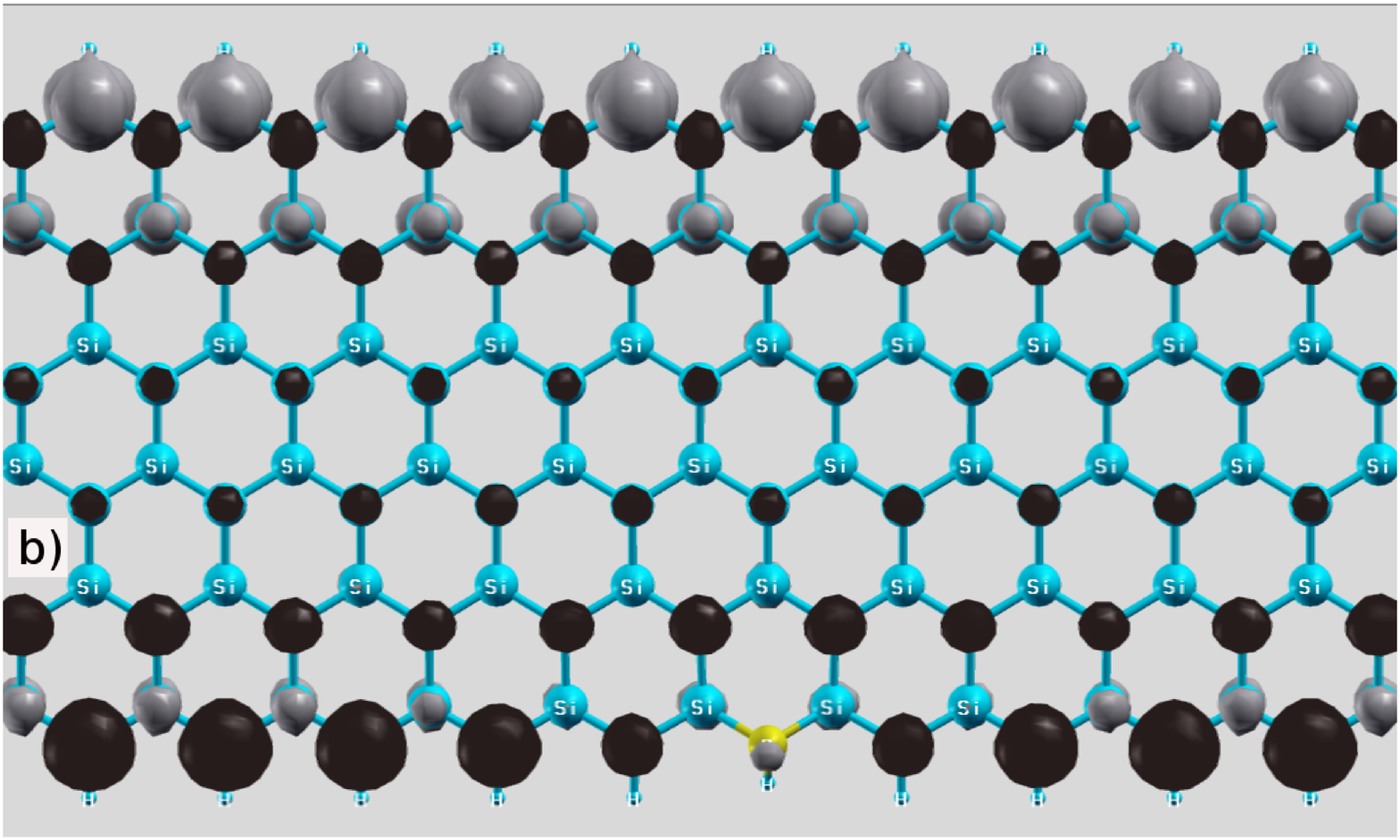}} \\
      \resizebox{42mm}{!}{\includegraphics[angle=0]{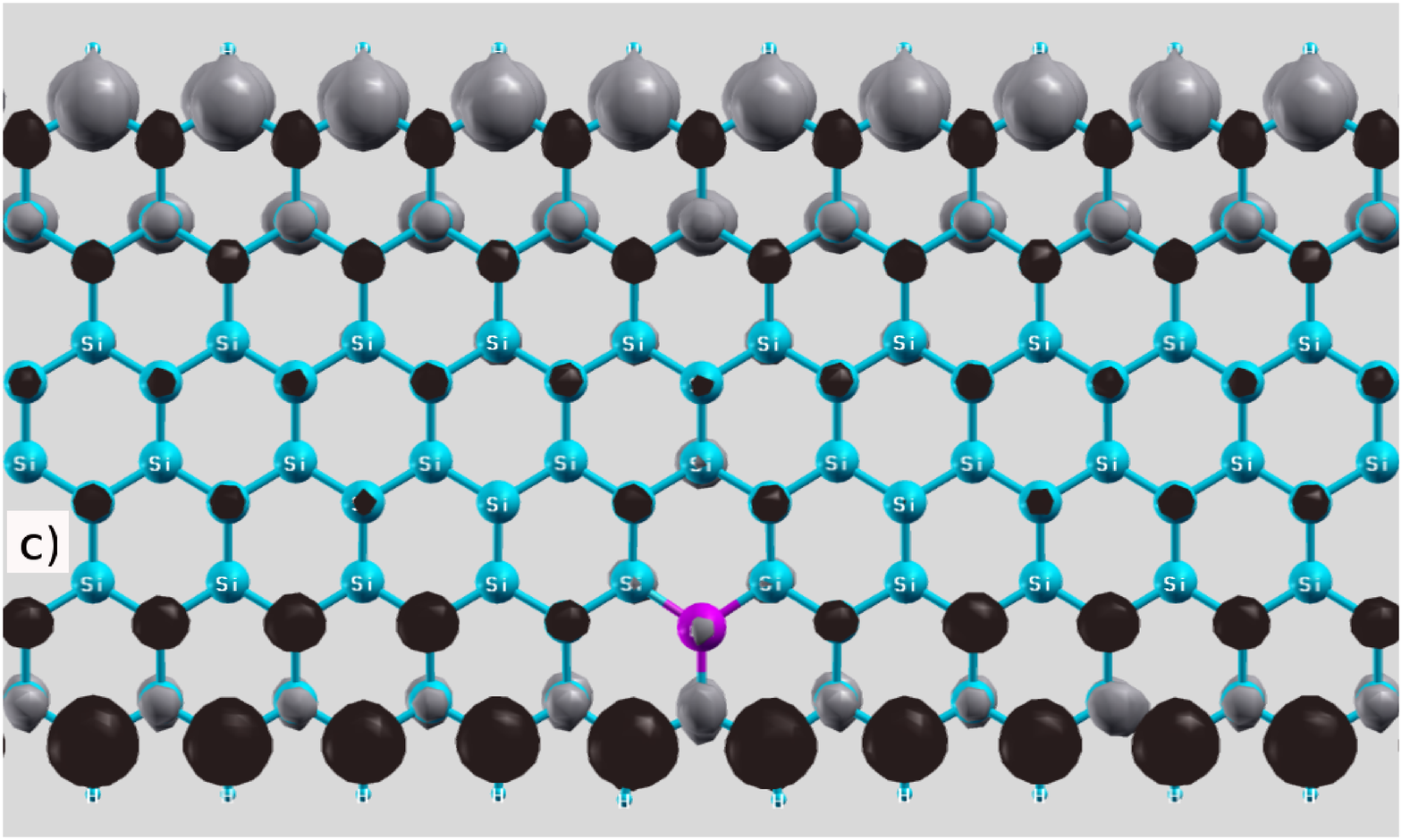}} &
      \resizebox{42mm}{!}{\includegraphics[angle=0]{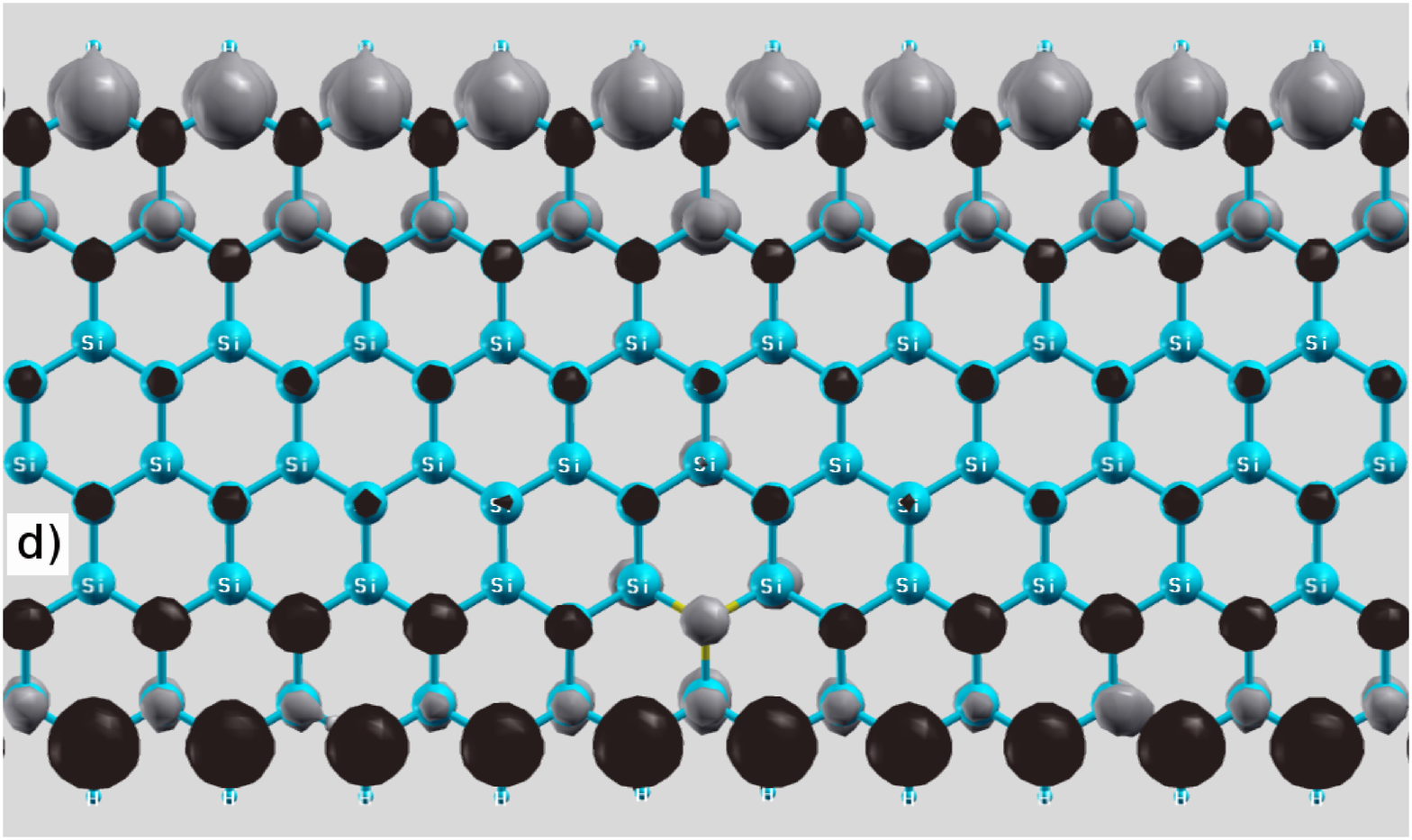}} \\
      \resizebox{42mm}{!}{\includegraphics[angle=0]{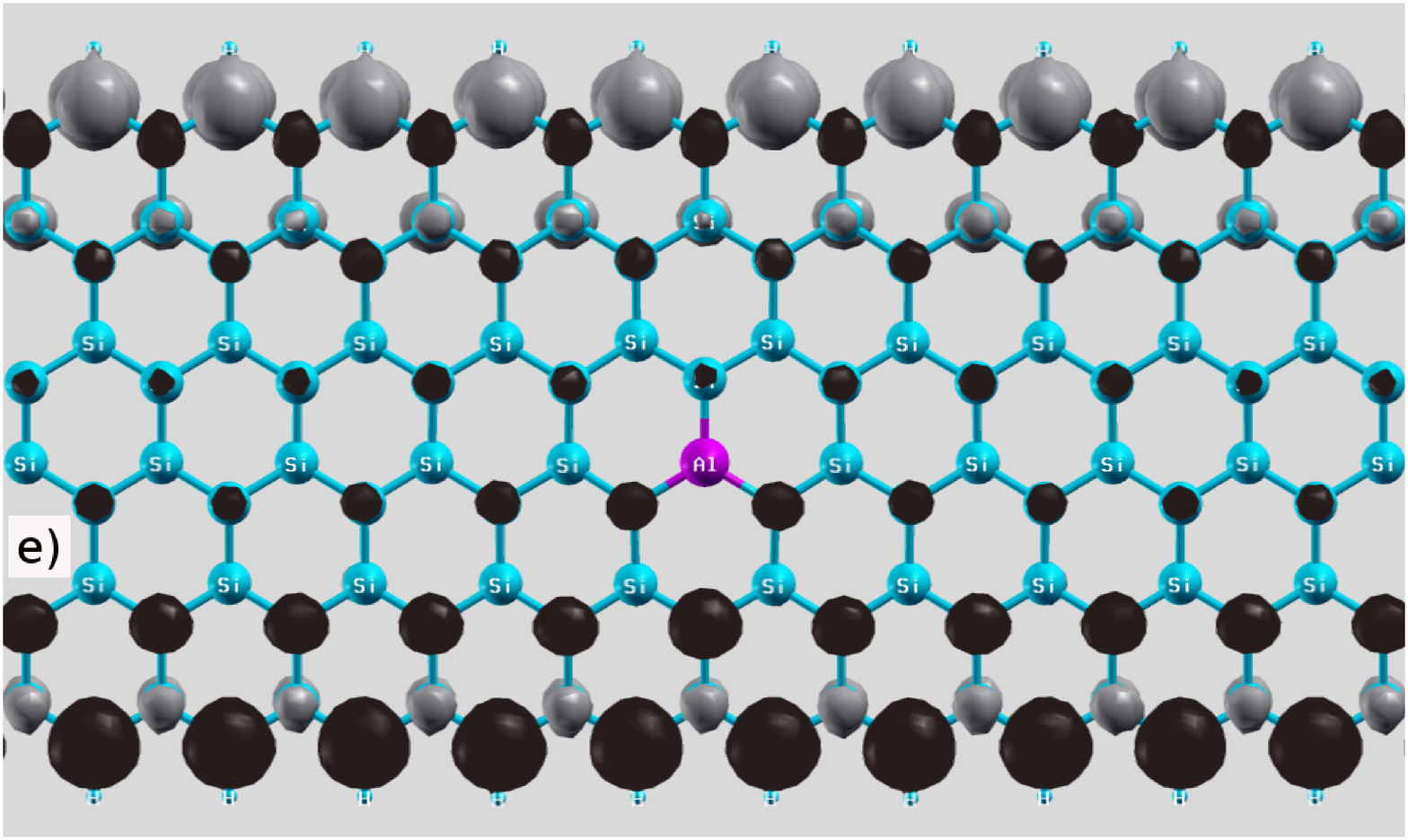}} &
      \resizebox{42mm}{!}{\includegraphics[angle=0]{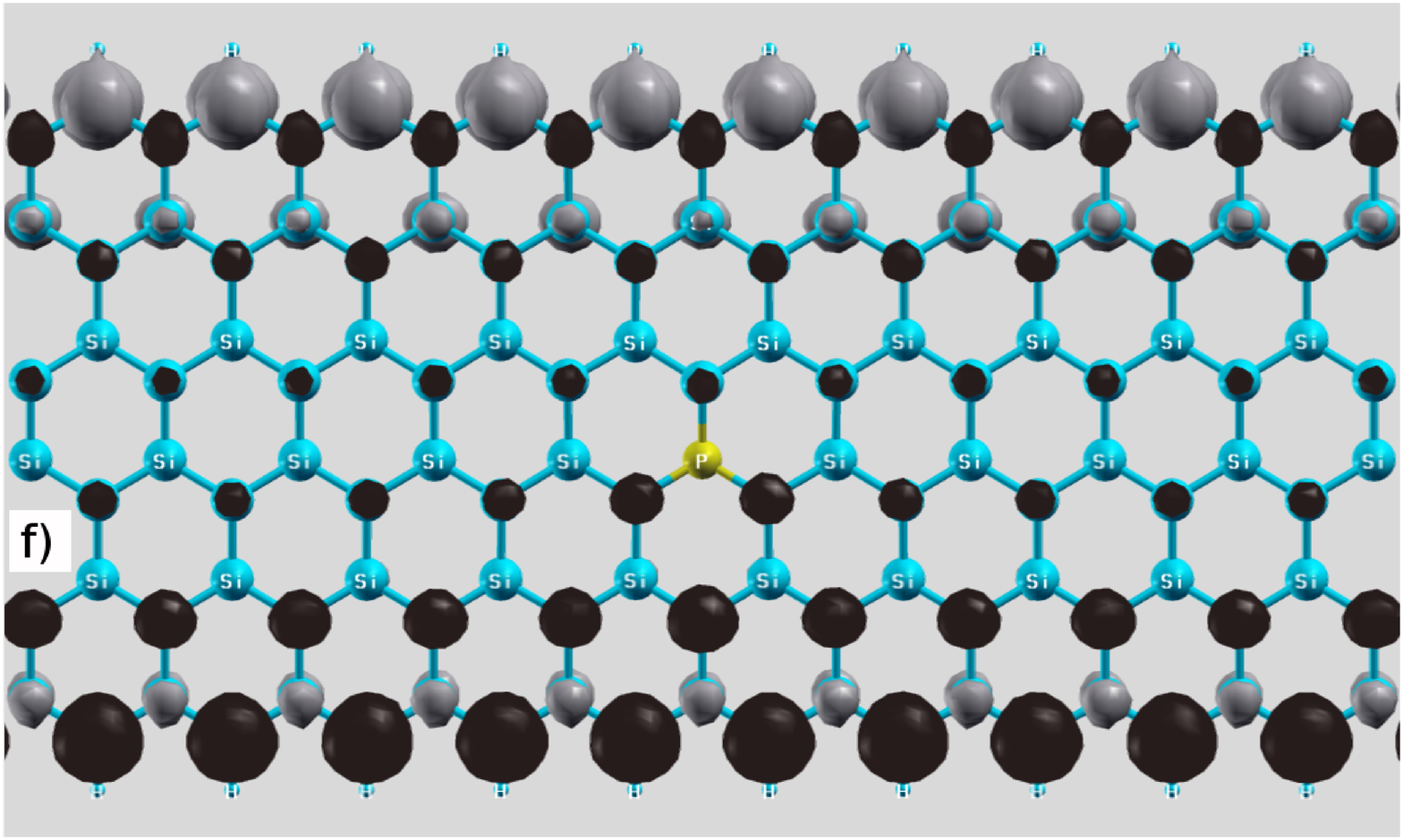}} \\
    \end{tabular}
    \caption{(Color online) Spin density in the low energy state, calculated within GGA approximation for zSiNRs with $N=6$
             for Al (left panel) and P (right panel) impurities in the  PE (a,b), PM (c,d) and PC (e,f) configurations.
             Black and gray dots represent magnetic moments of opposite directions.}
    \label{fig1}
  \end{center}
\end{figure}

\begin{figure}[ht]
  \begin{center}
    \begin{tabular}{cc}
      \resizebox{42mm}{!}{\includegraphics[angle=0]{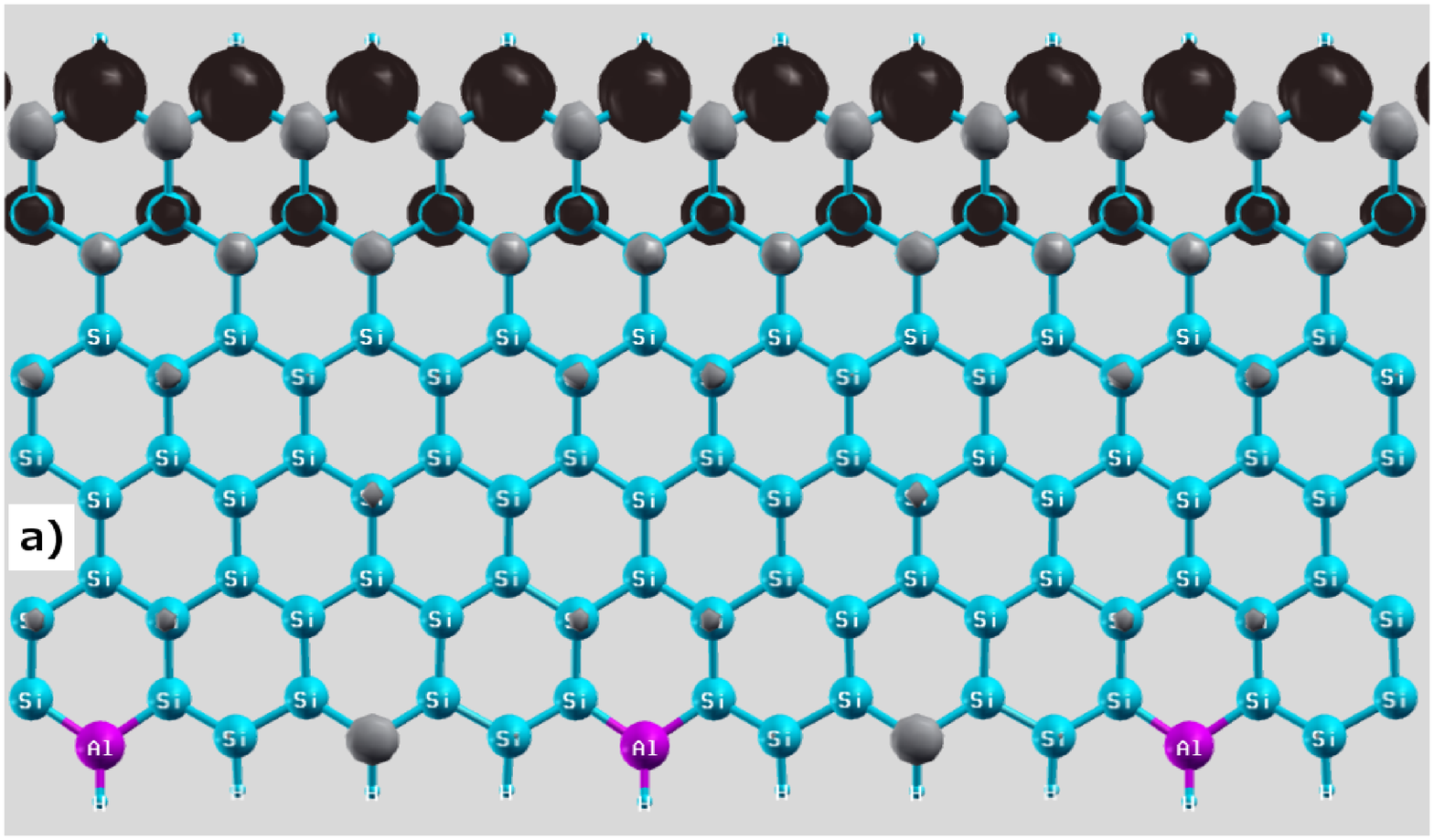}} &
      \resizebox{42mm}{!}{\includegraphics[angle=0]{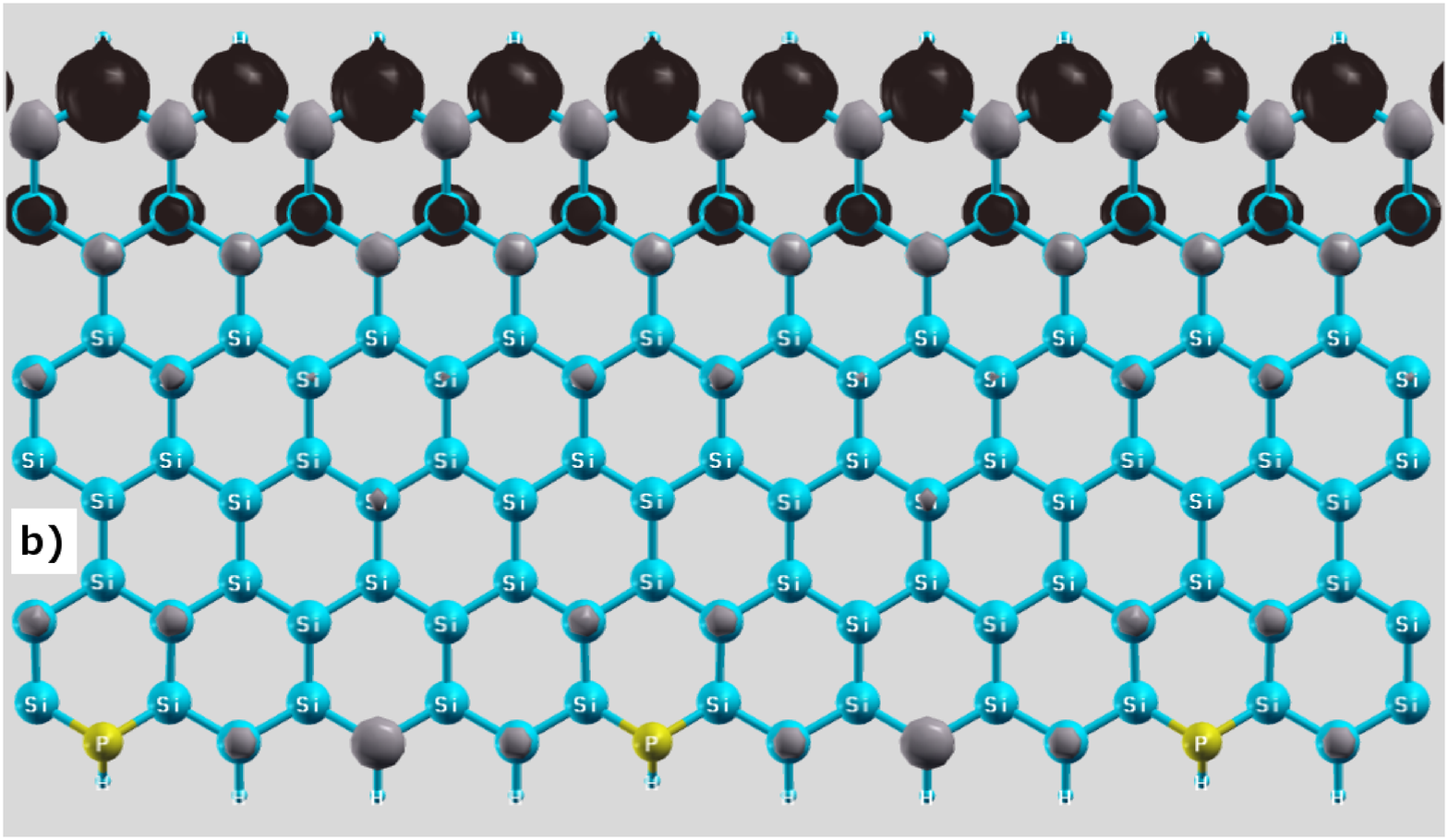}} \\
    \end{tabular}
    \caption{(Color online) Spin density in the one-edge FM state, calculated within GGA approximation for zSiNRs with $N=6$ for Al (left) and P (right) impurities in the  PE configuration.
              Note, the density of impurity atoms is here larger than in Fig.~\ref{fig1}. Meaning of the black and gray dots same as in Fig.~\ref{fig1}.}
    \label{fig0}
  \end{center}
\end{figure}

In Fig.~\ref{fig1} we show spin density for zSiNRs in the ground state (referred to in the following as the low energy state) for the three different impurity configurations, i.e. PE, PC and PM ones.
In the presence of non-magnetic impurity atoms (Al, P), magnetic moments at the two nanoribbon edges do not fully compensate each other. Moreover,
some small moments are also localized on inner atoms (see Fig.~\ref{fig1}). As a result, small
net magnetization can be observed in the low energy state.
Our calculations also show, that when the impurity concentration at the edge (PE configuration) increases, the
low-energy state becomes ferromagnetic, with magnetic moments located only at the impurity-free
edge of the nanoribbon (see Fig.~\ref{fig0}). This happens when the distance between the edge impurity
atoms is smaller than 19 $\angstr$, which is in agreement with Refs~\cite{28,29}.
Thus, the low-energy state is antiferromagnetic for pristine nanoribbons, generally ferrimagnetic in nanoribbons with impurities, and ferromagnetic for impurities located at one edge and of a sufficiently large concentration.
Since smaller impurity concentrations are easier to be achieved experimentally,
we have performed calculations for zSiNRs with longer distance between the impurities. More specifically,
the distance is equal to the size of the elementary cell shown in Fig.1. Note, Fig.2 displays more elementary cells, as the distance between impurities is there smaller than in Fig.~\ref{fig1}. To be more specific, the elementary cell shown in Fig.~\ref{fig1} is the system through which transmission is calculated, while the left and and right semi-infinite parts of the nanoribbon are treated as external electrodes.

\begin{figure*}[ht]
 \begin{center}
    \begin{tabular}{cc}
      \resizebox{70mm}{!}{\includegraphics[angle=270]{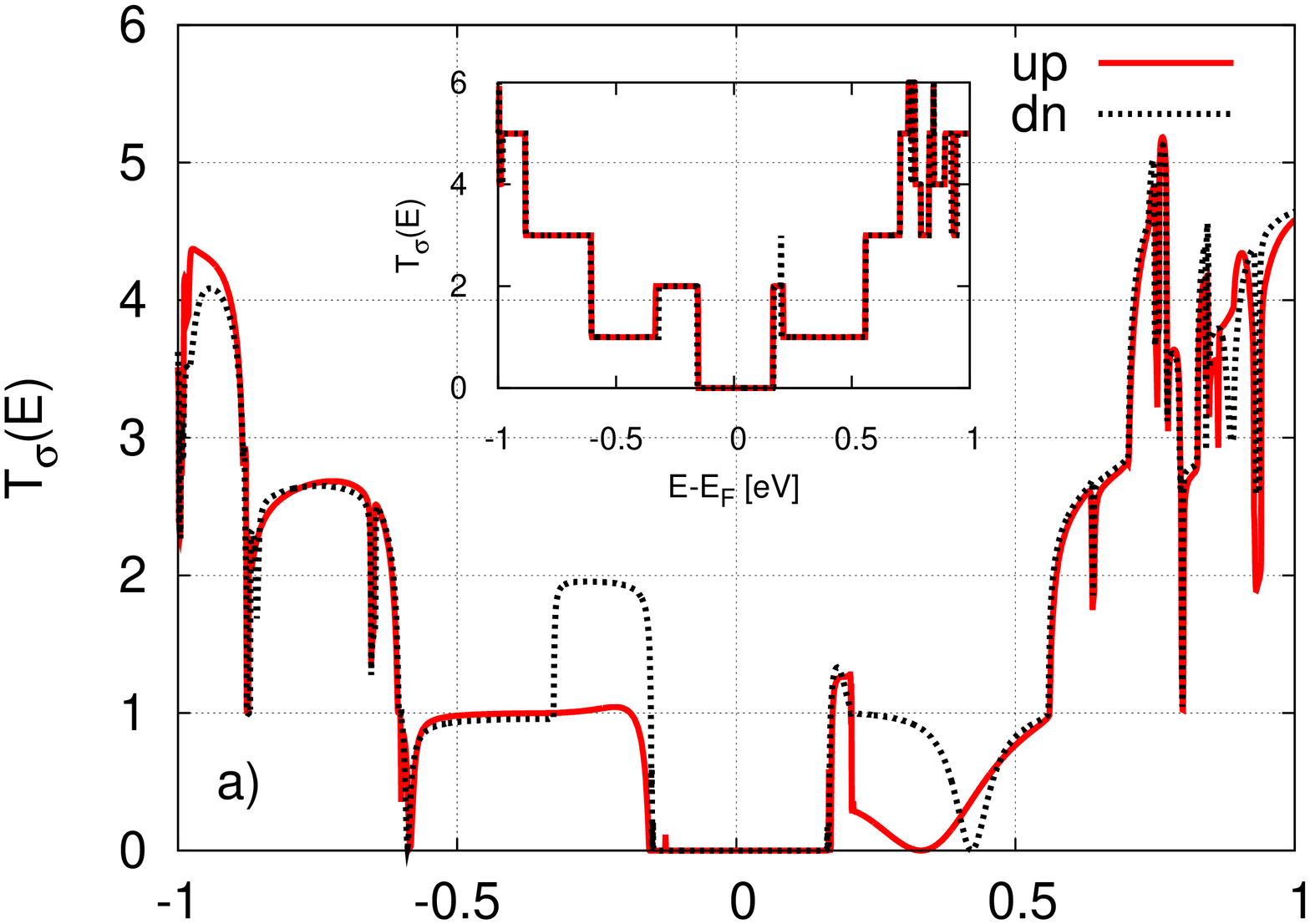}} &
      \resizebox{70mm}{!}{\includegraphics[angle=270]{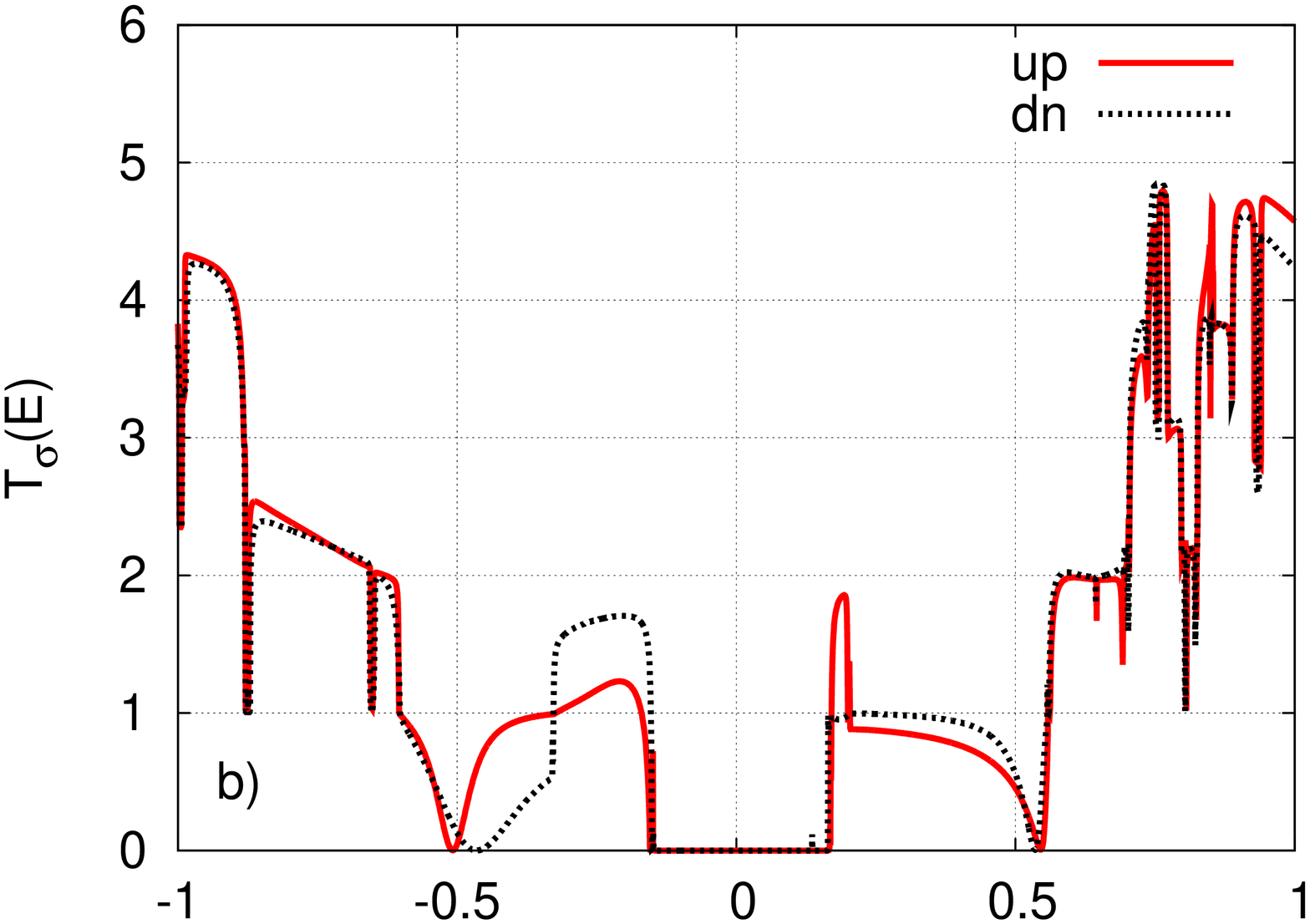}} \\
      \resizebox{70mm}{!}{\includegraphics[angle=270]{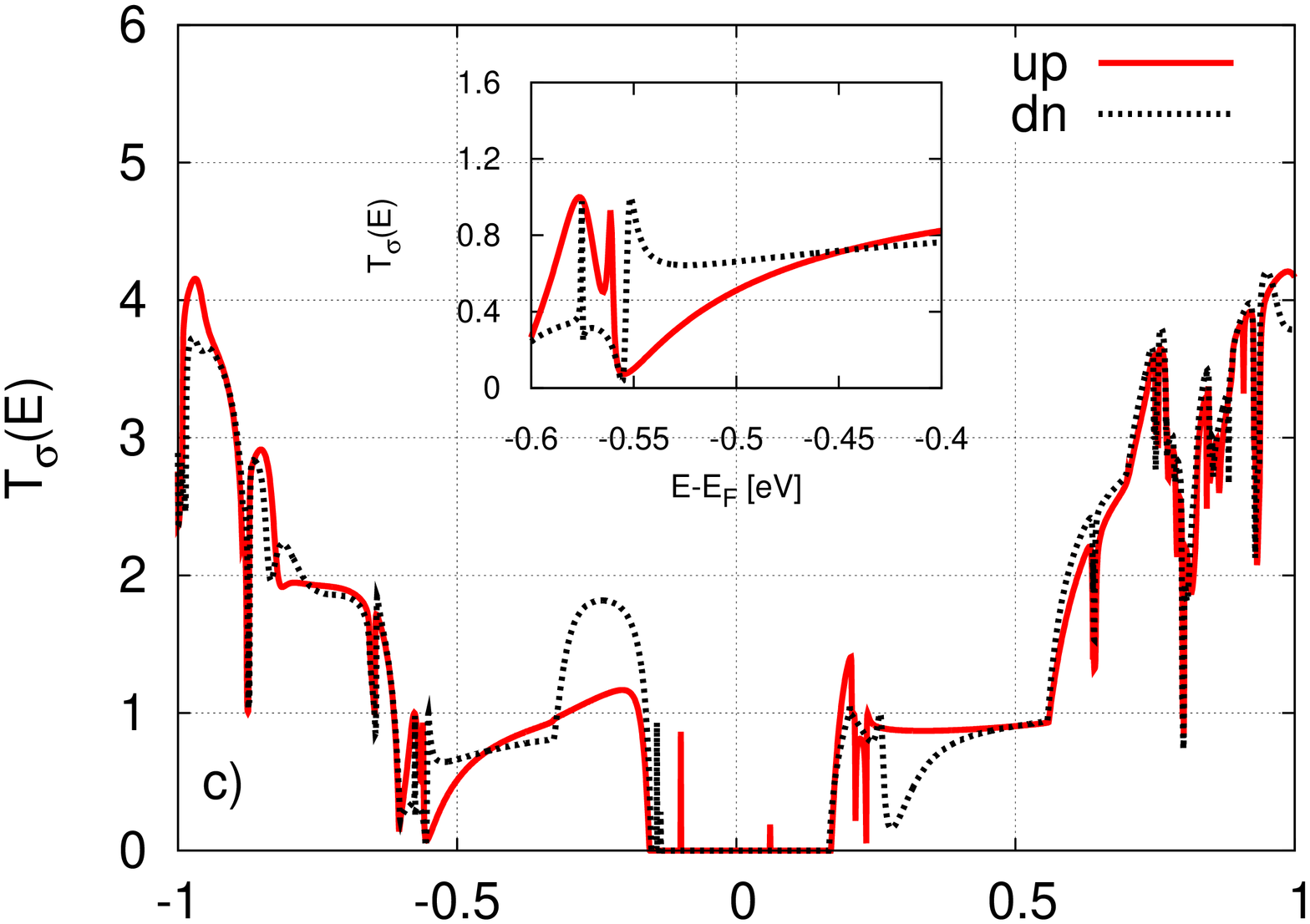}} &
      \resizebox{70mm}{!}{\includegraphics[angle=270]{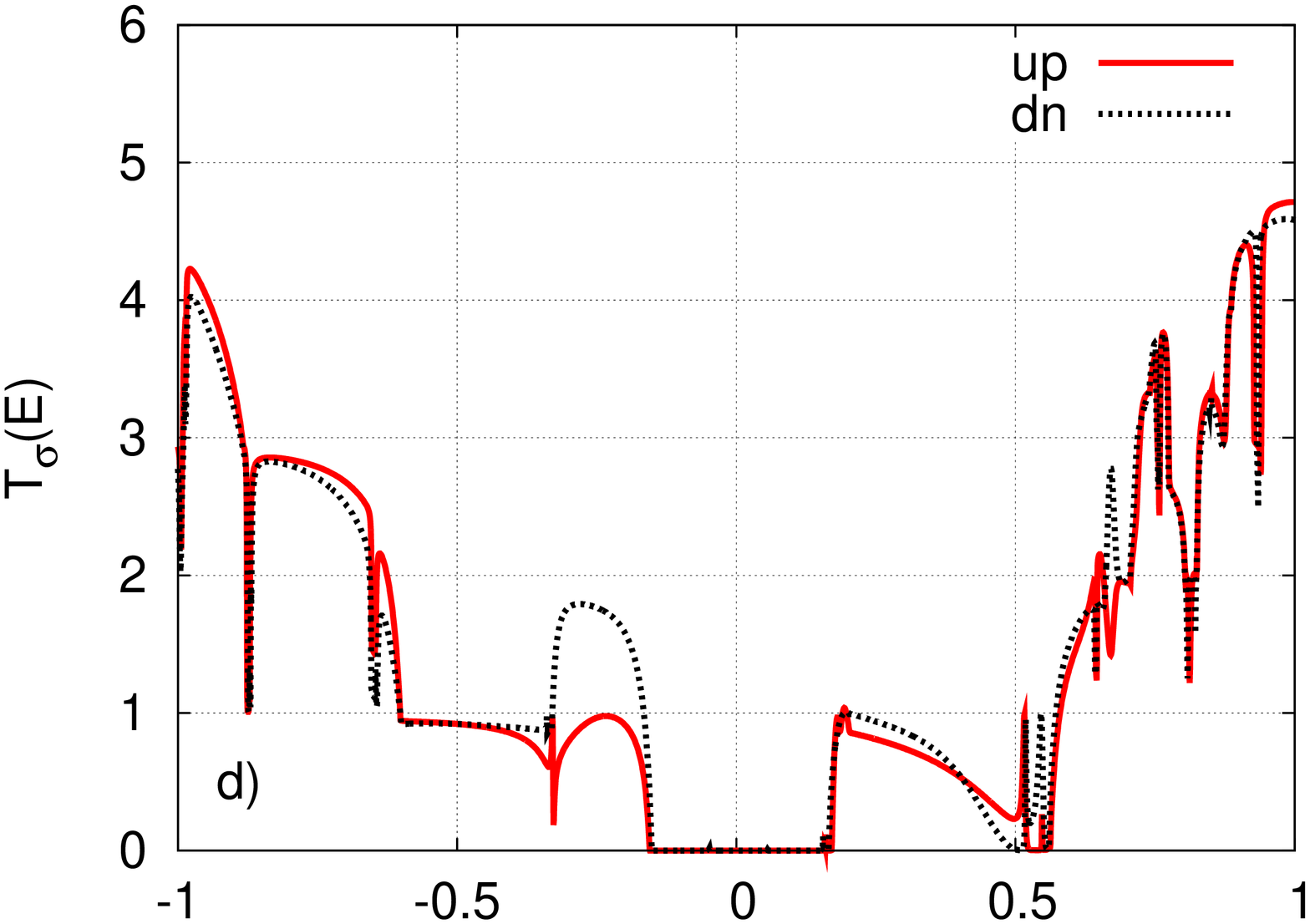}} \\
      \resizebox{70mm}{!}{\includegraphics[angle=270]{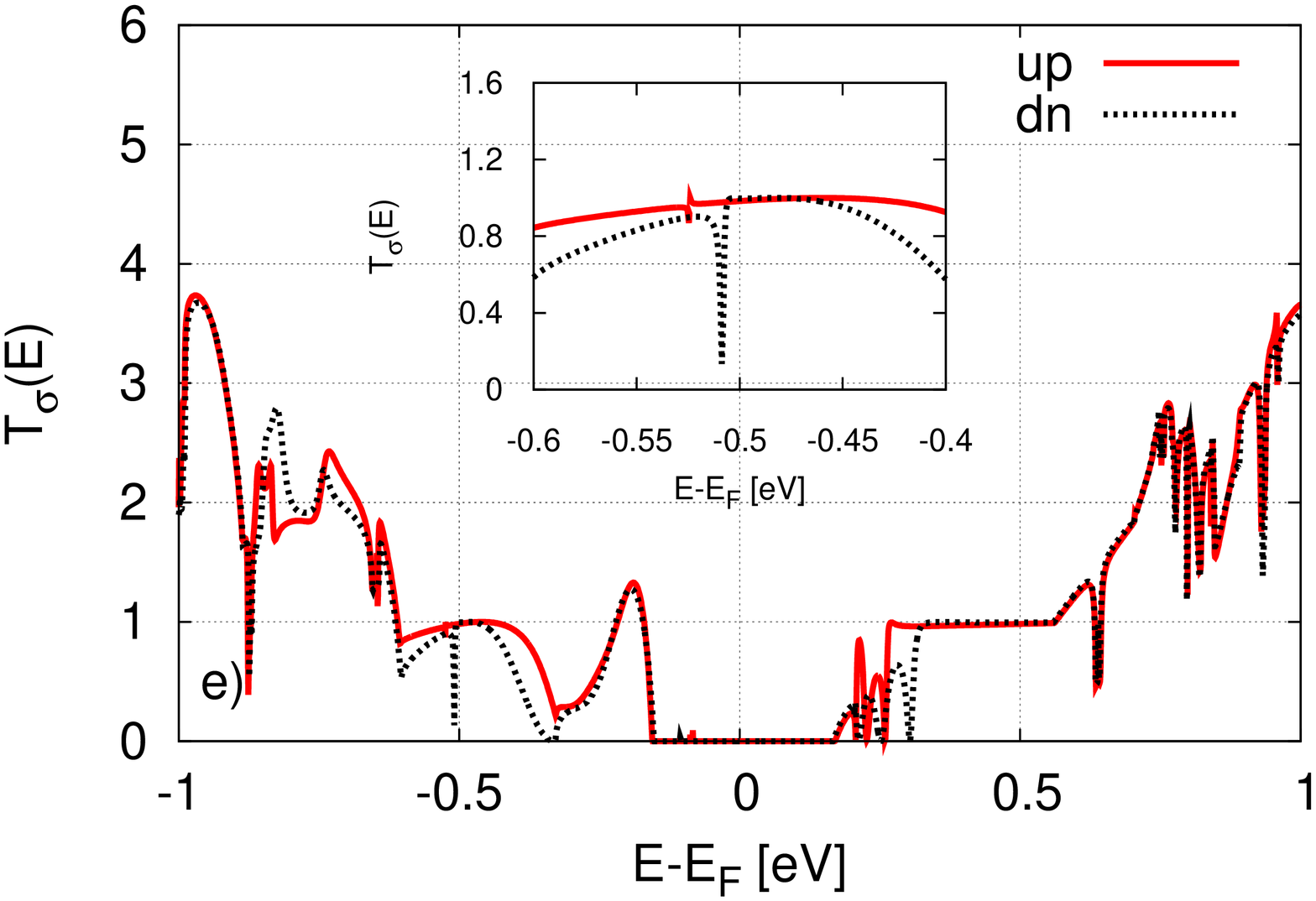}} &
      \resizebox{70mm}{!}{\includegraphics[angle=270]{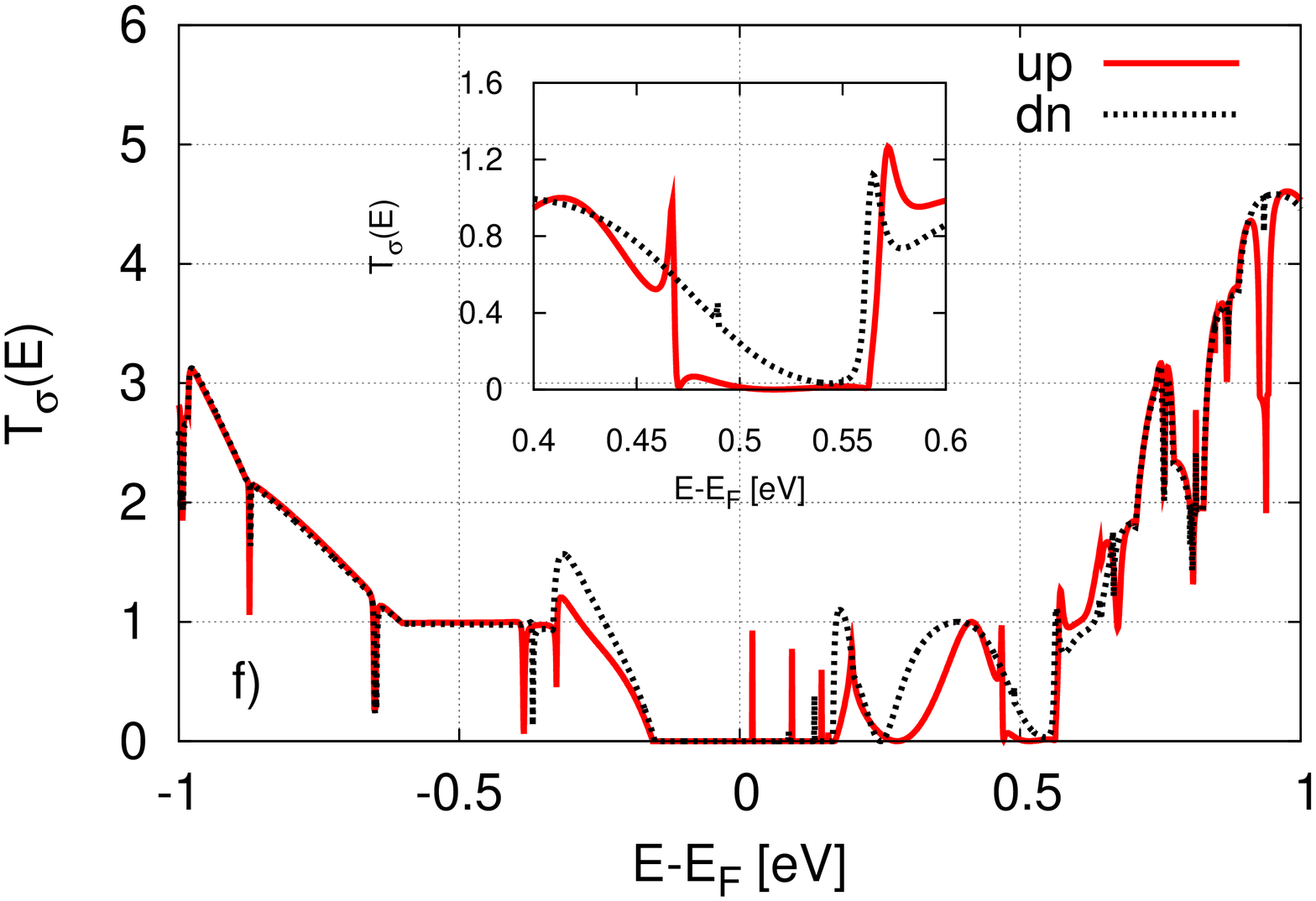}} \\
    \end{tabular}
    \caption{(Color online) Spin-dependent transmission in zSiNR ($N$=6) in the low-energy state as a function of energy, calculated within the GGA approximation
             for pristine nanoribbon (inset to a) and for Al (left panel) and P (right panel) impurities in the  PE (a,b), PM (c,d), and PC (e,f) configurations, respectively.
             Insets to Figs. c,e,f show transmission in a narrow energy range, where the Fano antiresonances are well resolved.}
    \label{fig2}
  \end{center}
\end{figure*}

\begin{figure}[ht]
  \begin{center}
    \begin{tabular}{cc}
      \resizebox{42mm}{!}{\includegraphics[angle=0]{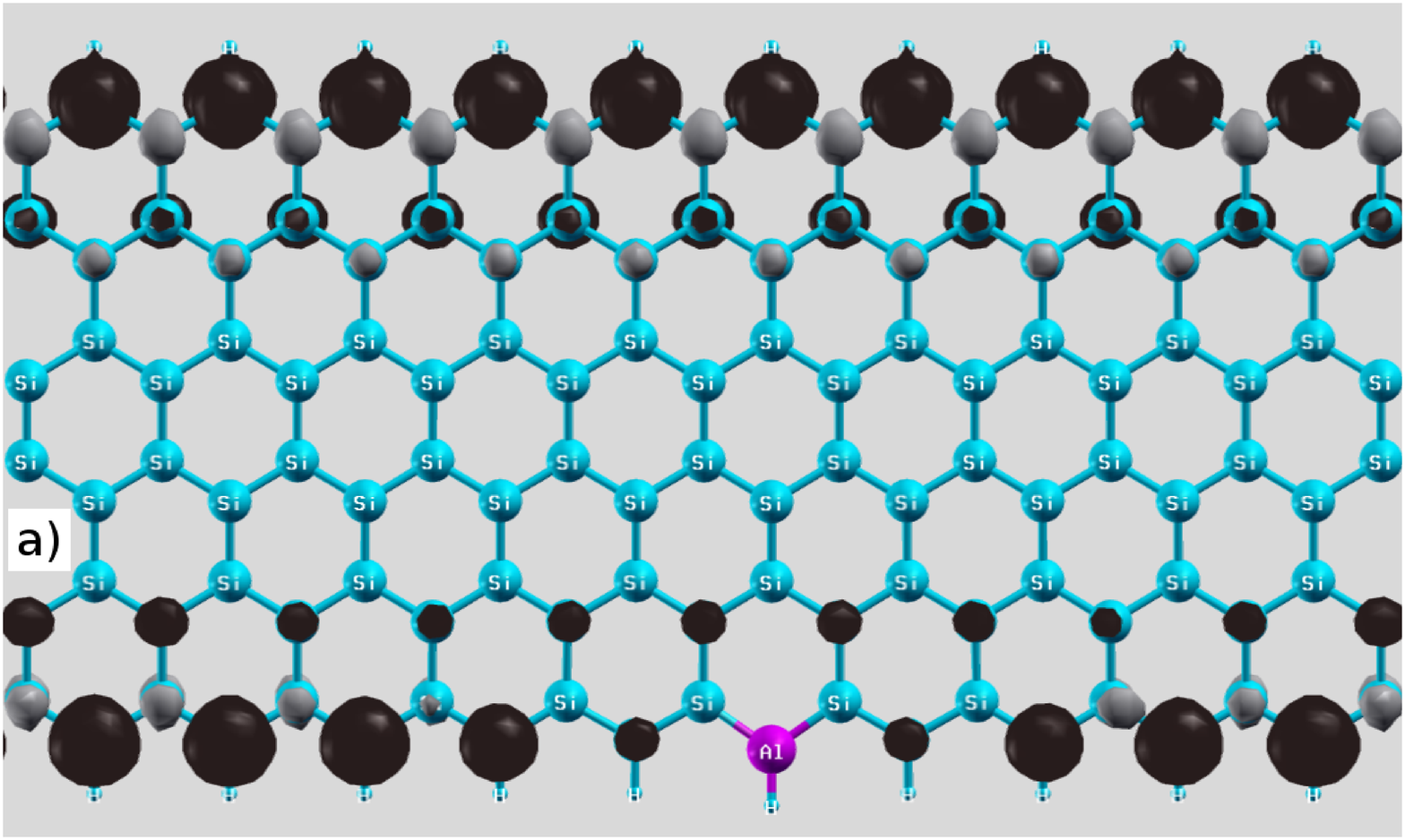}} &
      \resizebox{42mm}{!}{\includegraphics[angle=0]{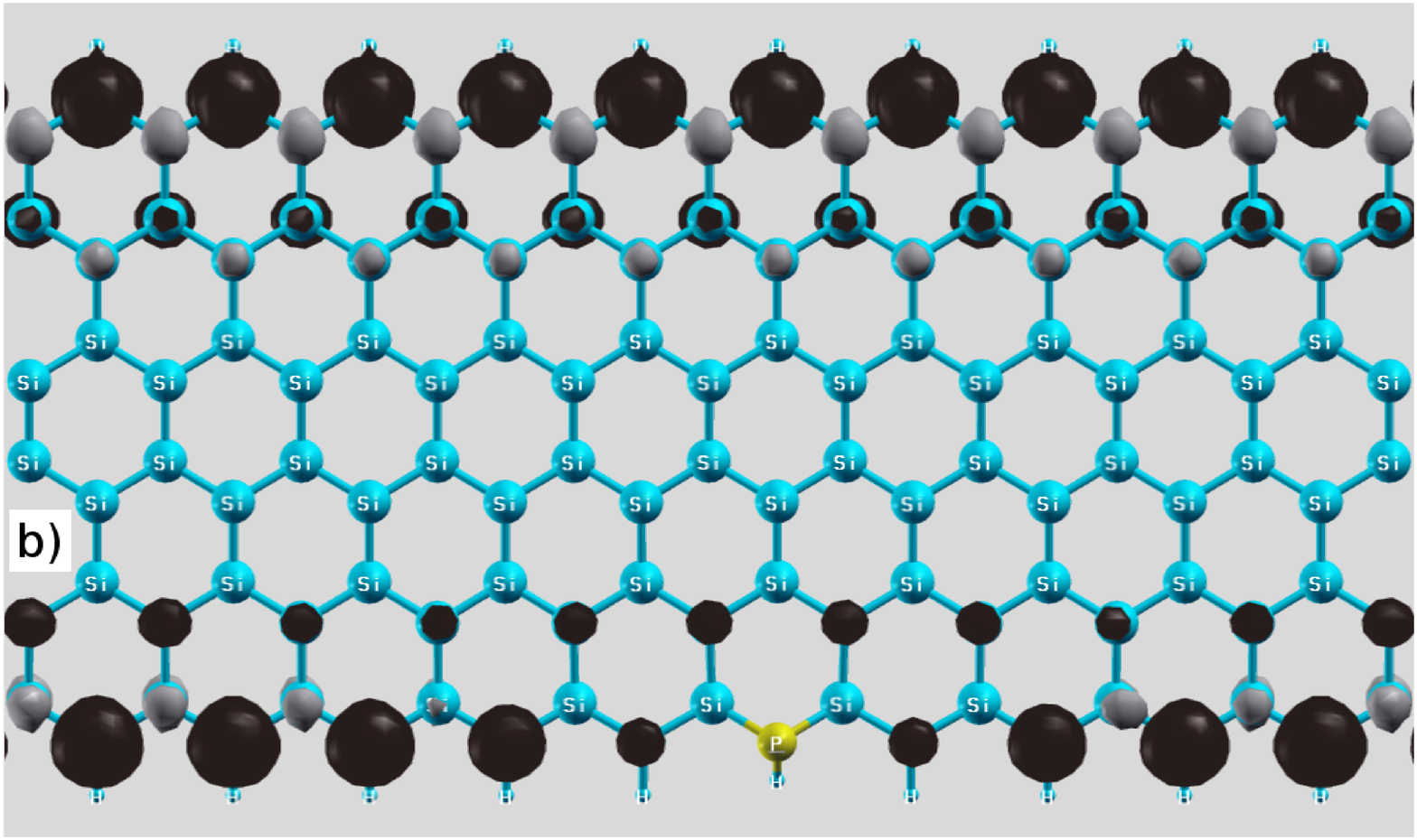}} \\
      \resizebox{42mm}{!}{\includegraphics[angle=0]{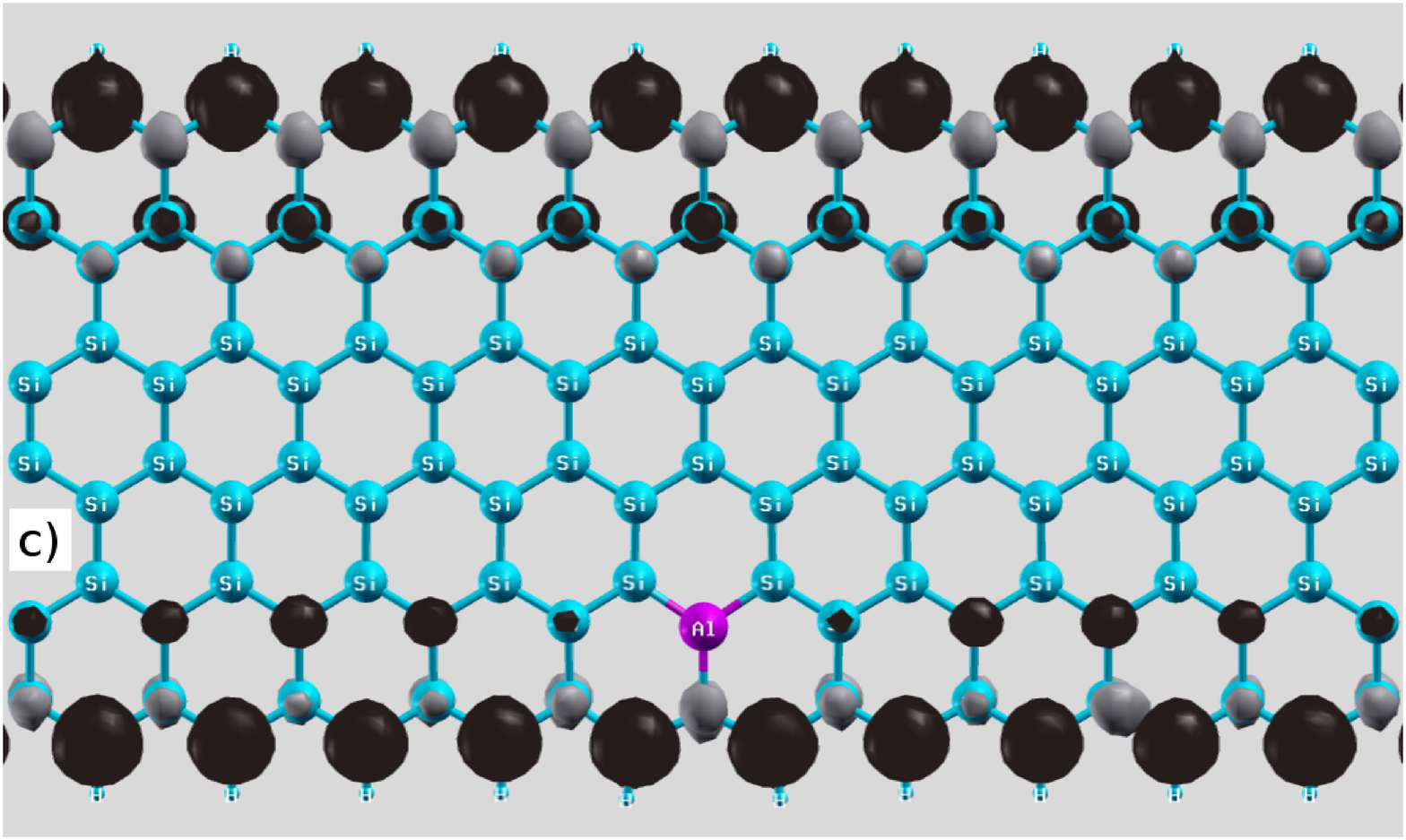}} &
      \resizebox{42mm}{!}{\includegraphics[angle=0]{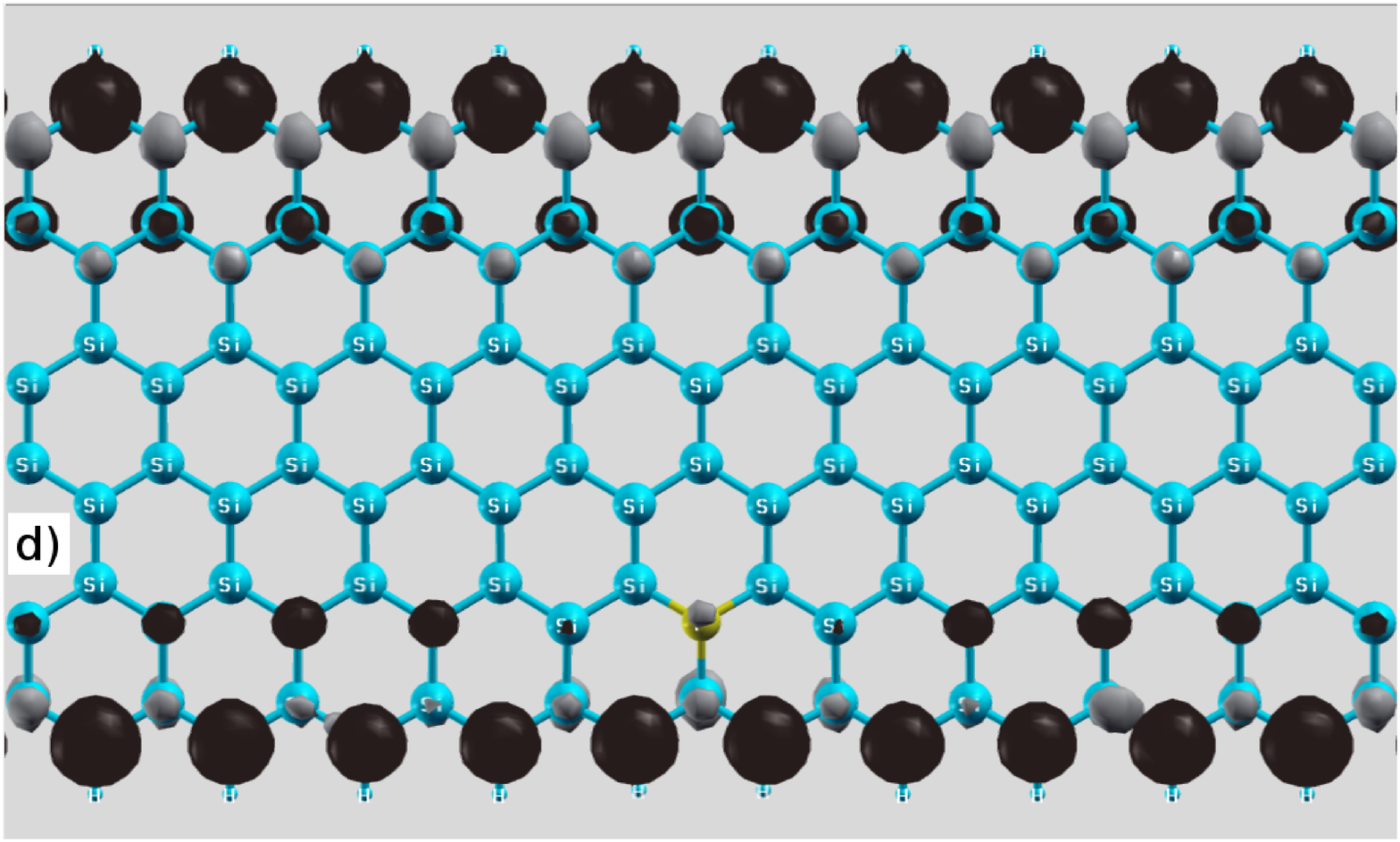}} \\
      \resizebox{42mm}{!}{\includegraphics[angle=0]{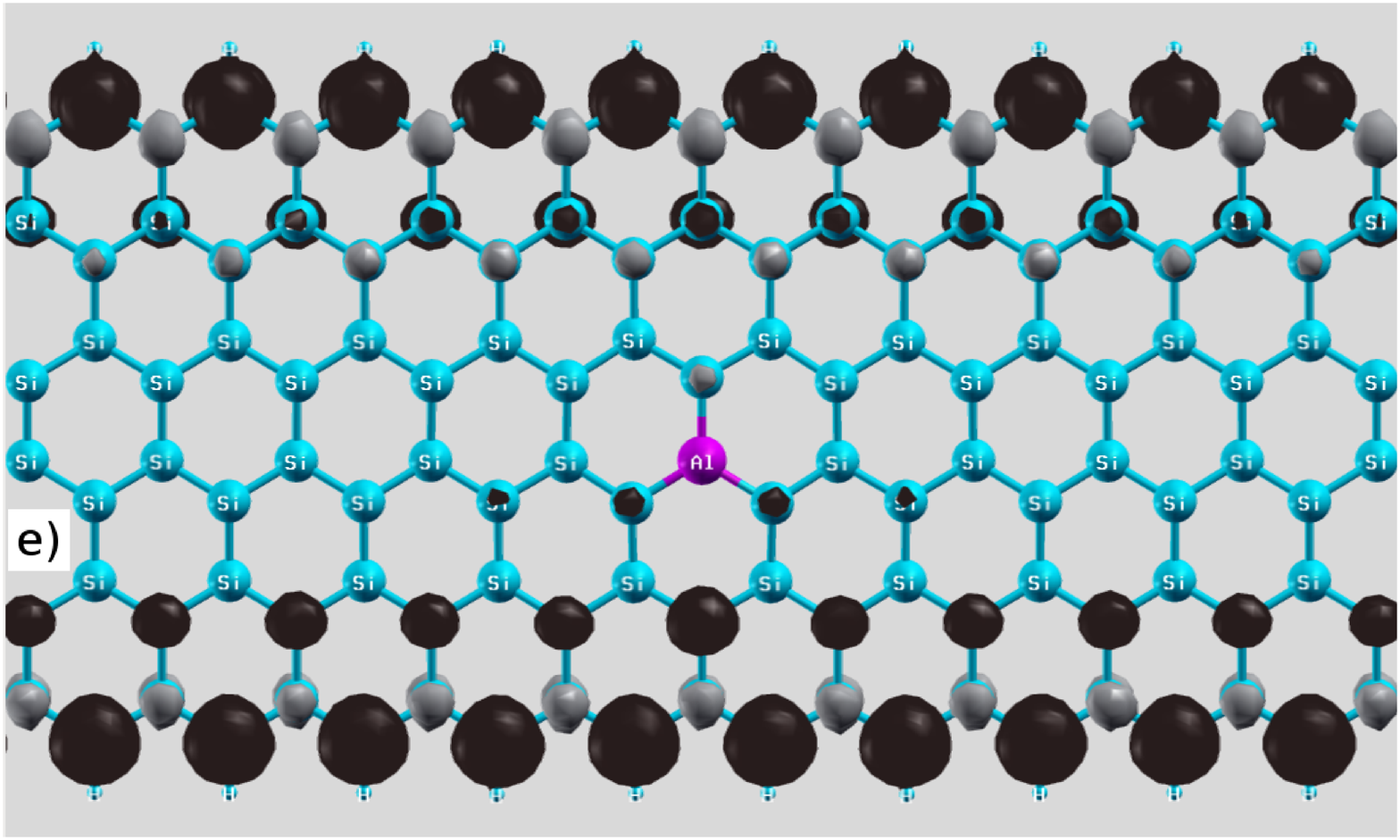}} &
      \resizebox{42mm}{!}{\includegraphics[angle=0]{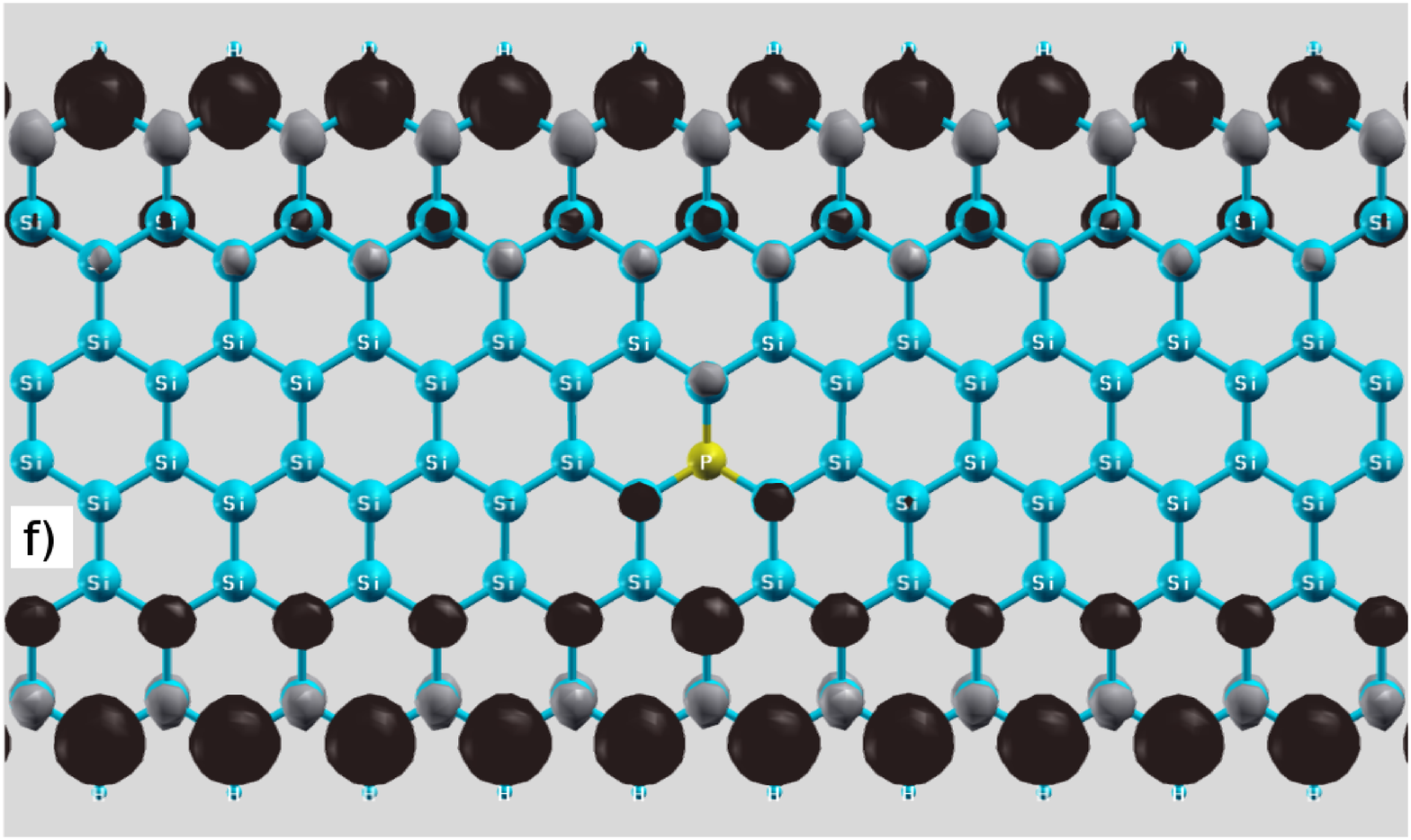}} \\
    \end{tabular}
    \caption{(Color online) Spin density in the FM state, calculated within GGA approximation for zSiNRs ($N$=6)
             for Al (left panel) and P (right panel) impurities in the PE (a,b), PM (c,d) and PC (e,f) configurations, respectively.
             Meaning of the black and gray dots same as in Fig.~\ref{fig1}.}
    \label{fig3}
  \end{center}
\end{figure}

Transmission function $T_{\sigma}(E)$ corresponding to the nanoribbons shown in Fig.~\ref{fig1} is presented in Fig.~\ref{fig2}
for both spin orientations. The transmission is shown there as a function of energy measured from the corresponding Fermi energy $E_F$.
One can note that the energy gap at the Fermi level, which exists in pristine
zSiNRs in the AFM state (inset to Fig.~\ref{fig2}a), survives also in the presence of impurity atoms. However, transmission depends now on the
spin orientation because magnetic moments of the two edges do not fully compensate each other in the presence of impurities.
Moreover, $T_{\sigma}(E)$ depends on the  type
and position of impurities. The transmission is strongly modified near the edges of the energy gap.
The impurity atoms can lead to states localized in the gap, which give narrow
peaks in the transmission function -- especially well visible for Al atoms in the PM  configuration (below the Fermi level) and in the PC configuration of
P impurities (above the Fermi level). Apart from the wide
gap near the Fermi energy $E_F$, additional gaps appear in the spectrum at other energies. A wide gap opens
above the Fermi level in the spin-up channel for Al impurities
localized at the nanoribbon edge, and a similar gap appears  below the Fermi level in the spin-down channel for P impurities (Fig.~\ref{fig2}).
It is interesting to note that spin-down electrons in the former case and spin-up holes in the latter case are
less influenced by the impurities, so the corresponding gaps are much narrower.
The transmission exhibits also a number of antiresonance dips -- mainly for higher values of
$|E - E_{F}|$. In the presence of impurity atoms, quantum
interference can lead to Fano antiresonances. Especially interesting results are obtained
for the PC impurity configuration, where  typical and well defined Fano antiresonances
in nanoribbons with P impurity atoms appear in the transmission in the vicinity of  $E - E_{F}$ = 0.5 eV
(Fig.~\ref{fig2}f). Due to the destructive interference, a relatively wide gap appears in this
energy region,  with transmission close to
zero. On the other hand, in nanoribbons with Al impurity atoms in the PC configuration, the  Fano antiresonance can be observed near $E - E_{F}\approx -0.5$ eV as well as for
$E - E_{F} \approx  0.4$ eV. The interference effects are more pronounced for  spin-up carriers
in nanoribbons with P impurities, and spin-down carriers in nanoribbons with Al impurities, as considerable magnetic moments
are then localized on nearby Si atoms. Well defined Fano antiresonances can be also  visible for Al impurities
in the PM configuration (inset to Fig.~\ref{fig2}c)

\begin{figure*}[ht]
 \begin{center}
    \begin{tabular}{cc}
      \resizebox{70mm}{!}{\includegraphics[angle=270]{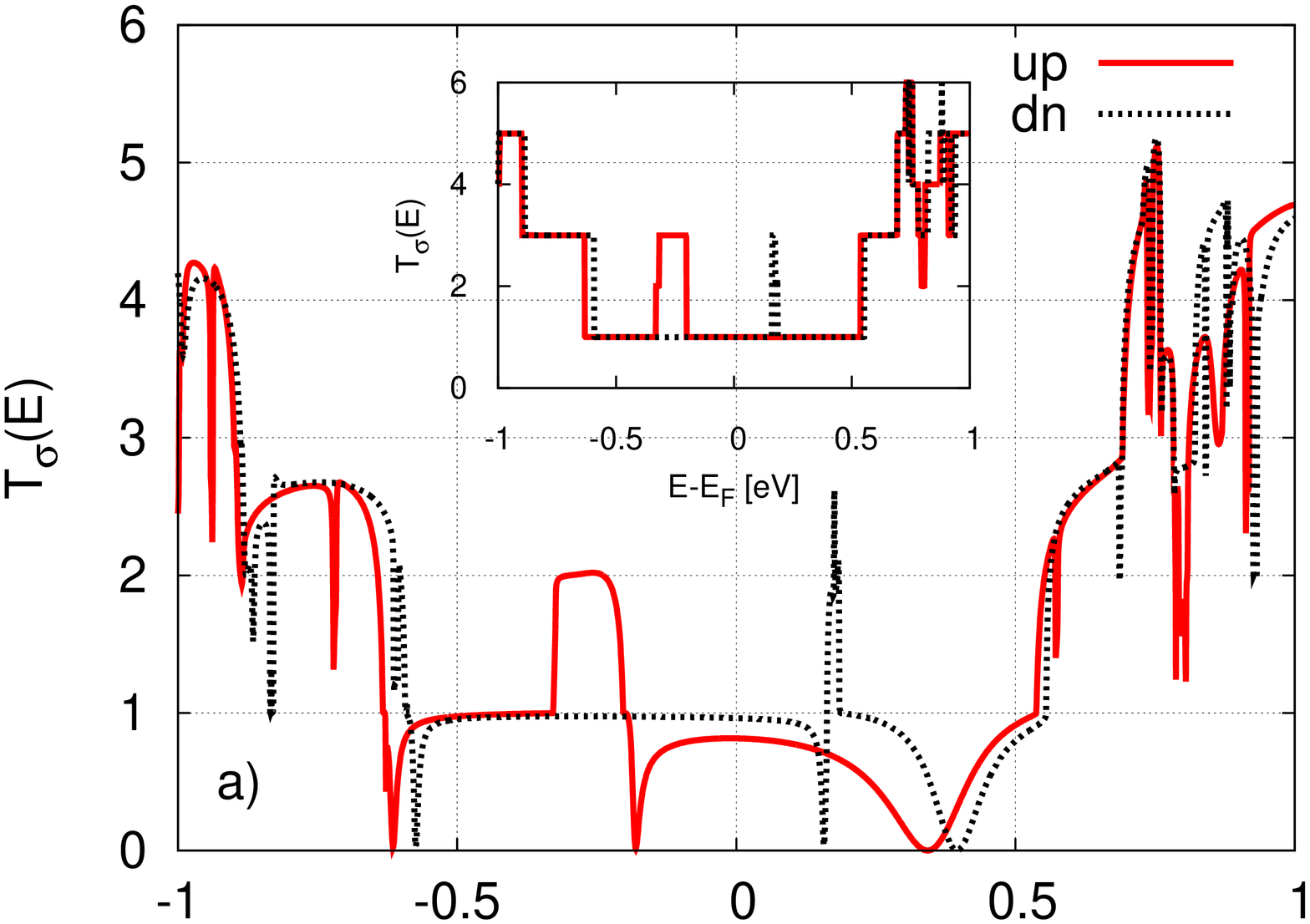}} &
      \resizebox{70mm}{!}{\includegraphics[angle=270]{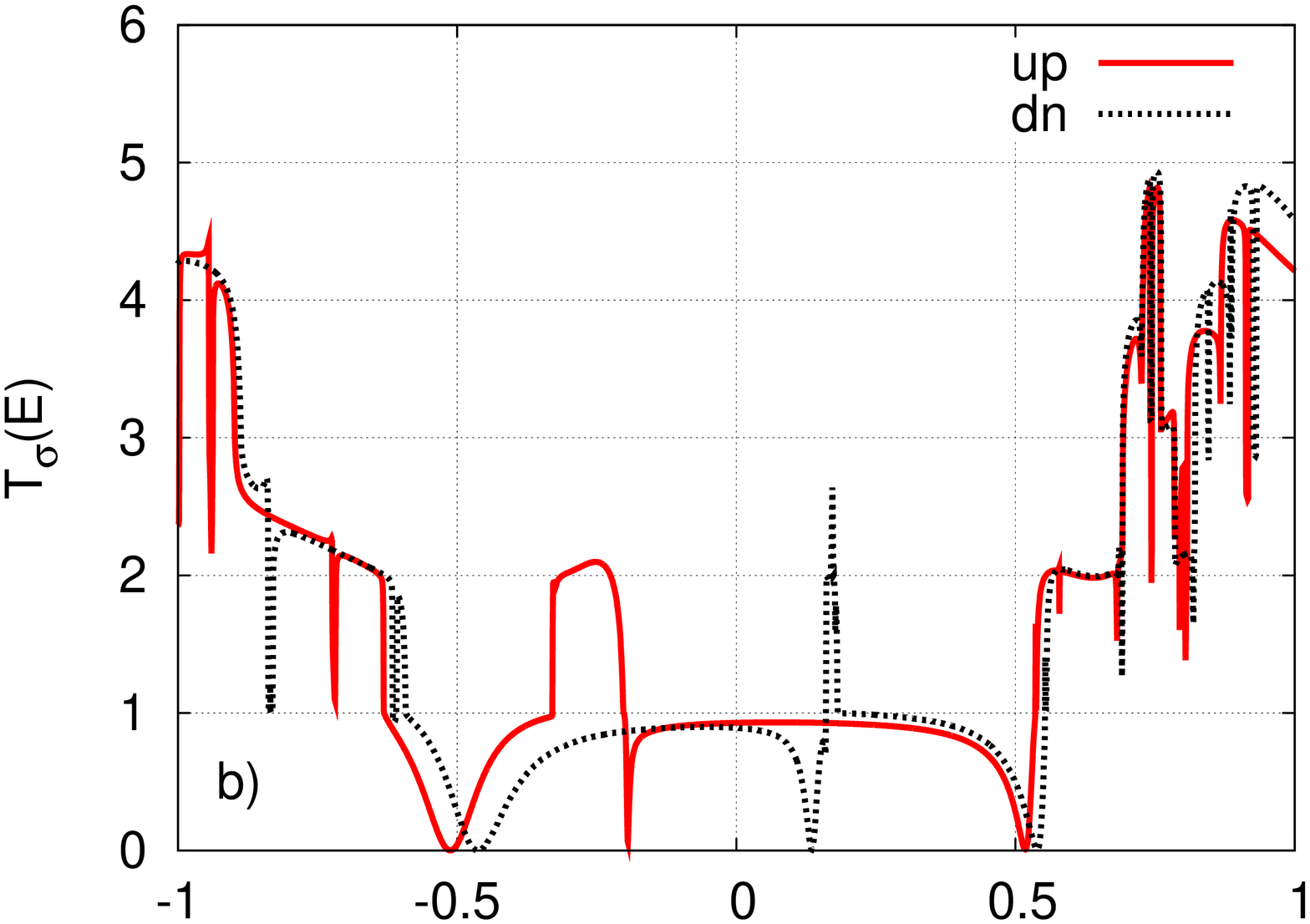}} \\
      \resizebox{70mm}{!}{\includegraphics[angle=270]{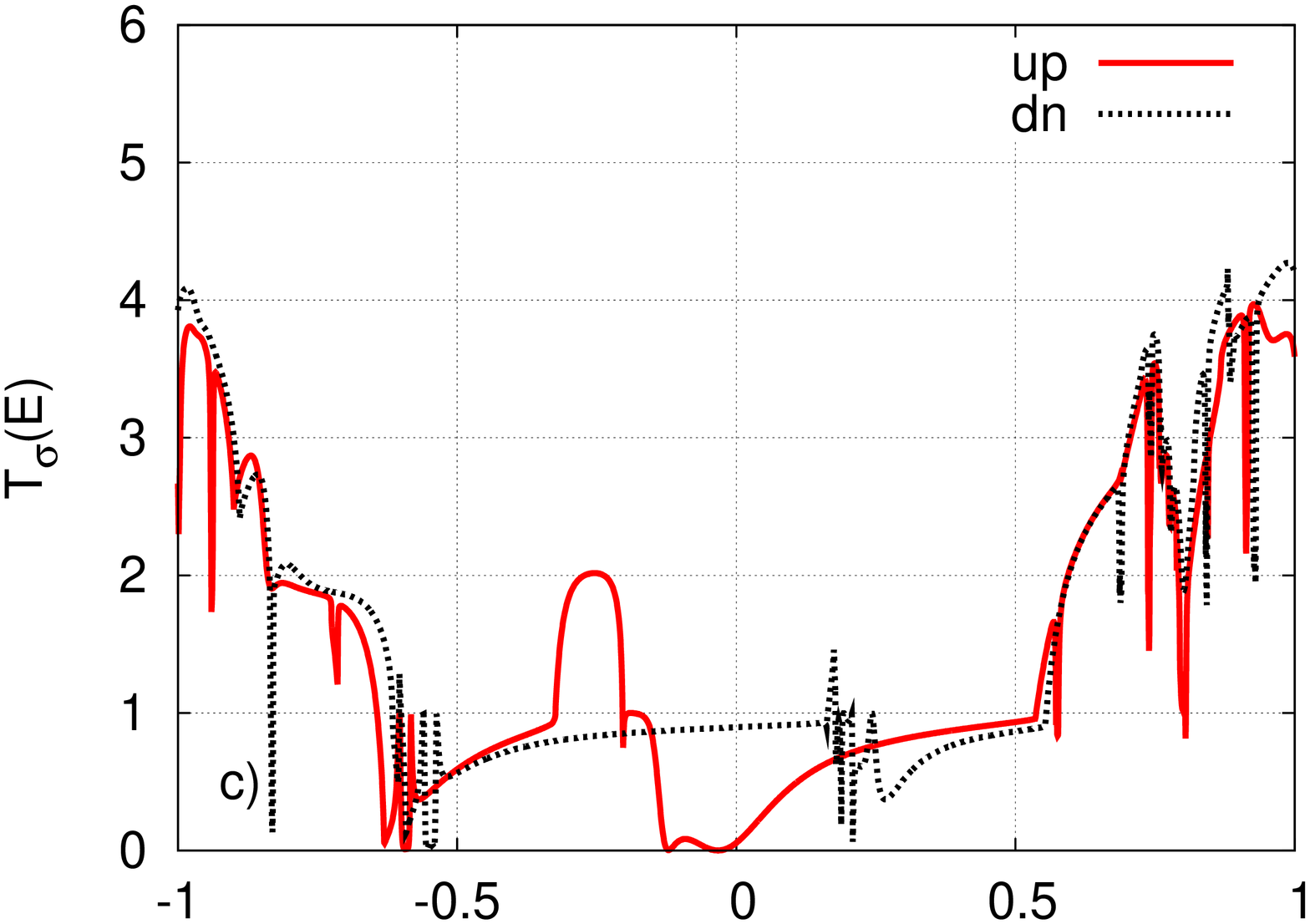}} &
      \resizebox{70mm}{!}{\includegraphics[angle=270]{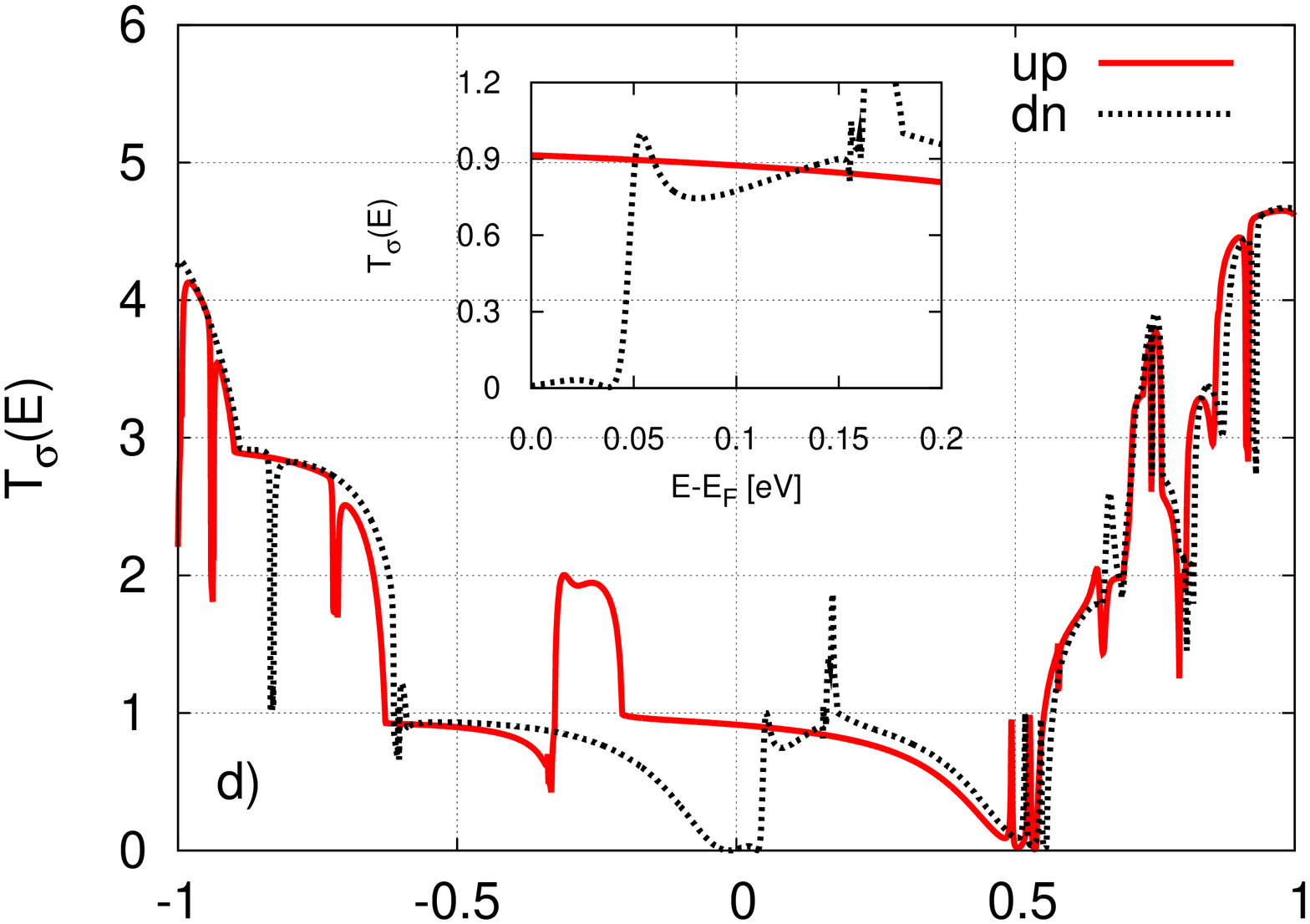}} \\
      \resizebox{70mm}{!}{\includegraphics[angle=270]{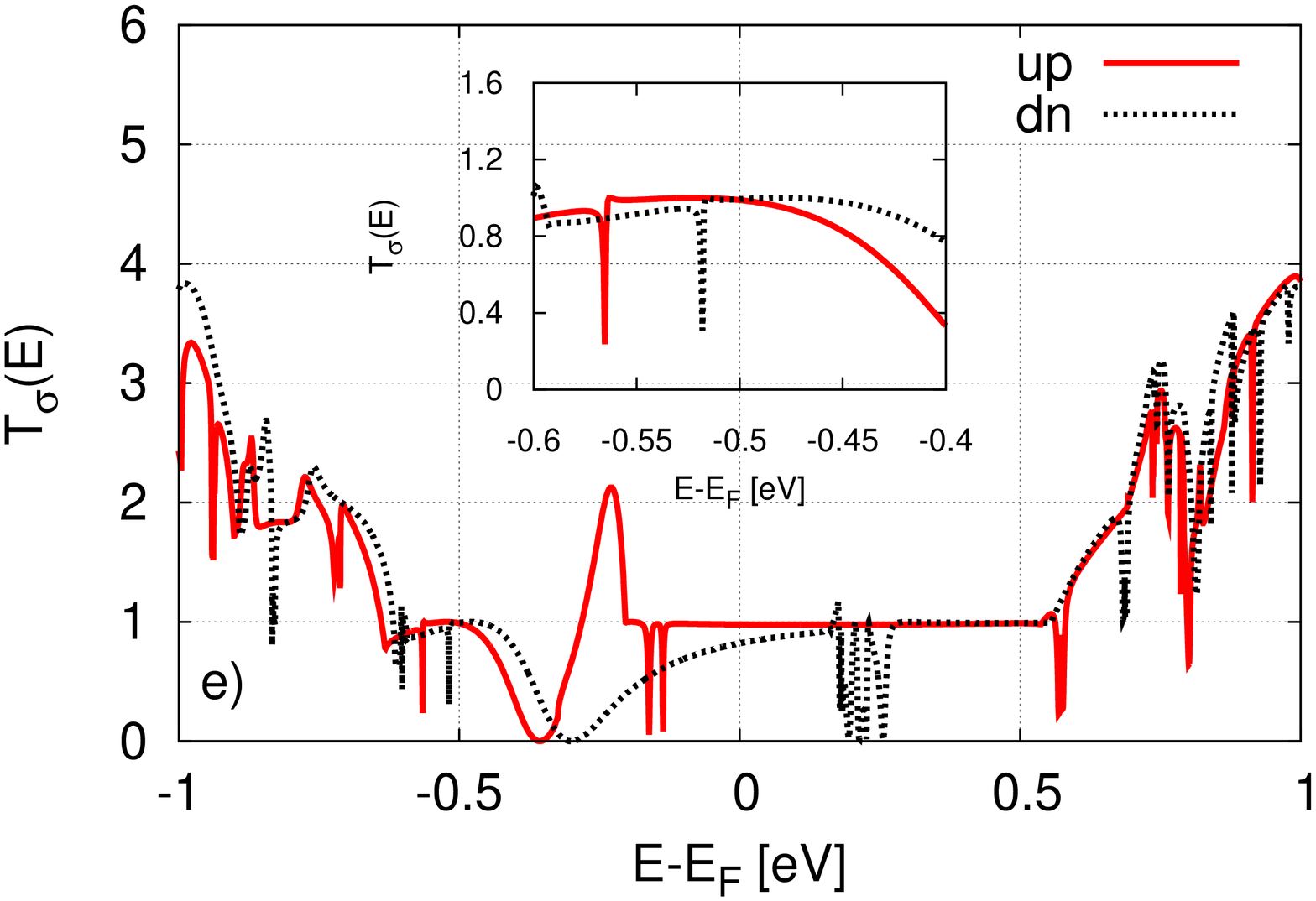}} &
      \resizebox{70mm}{!}{\includegraphics[angle=270]{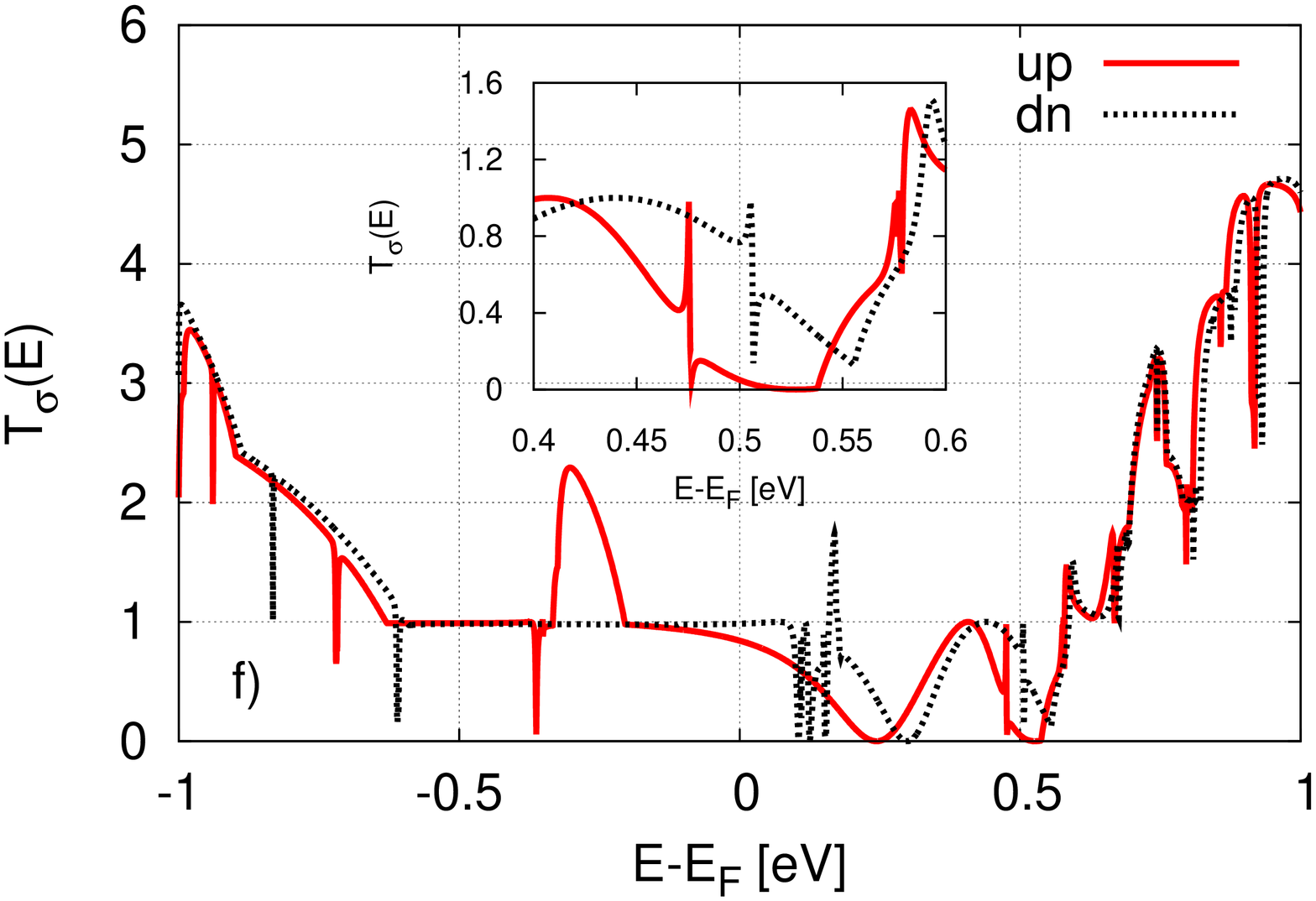}} \\
    \end{tabular}
    \caption{(Color online) Spin-dependent transmission for zSiNRs ($N$=6) in the FM state as a function of energy, calculated within the GGA approximation
             for pristine ribbon (inset to a) and for Al (left panel) and P (right panel) impurities in the PE (a,b), PM (c,d), PC (e,f) configurations.
             Insets to  e and f show the transmission in a narrow energy range, where the Fano antiresonances are well resolved.}
    \label{fig4}
  \end{center}
\end{figure*}

\subsection{Ferromagnetic state}

Consider now nanoribbons which display ferromagnetic ordering (FM state) in the absence as well as presence of impurities.
As already mentioned above, energy of such a state is only slightly
higher than that of the above discussed low-energy (ground) state. Thus, the FM state can be stabilized by an external magnetic
field or by other methods. Spin density for the three different configurations of impurity atoms is presented in Fig.~\ref{fig3}.
Note, magnetic moments localized at the two edges are not equal in the presence of impurities.

The corresponding spin-resolved transmission function is shown in  Fig.~\ref{fig4}.
Pristine ferromagnetic
zSiNRs exhibits typical metallic behavior with a constant transmission in the vicinity of the Fermi level, see the inset to Fig.~\ref{fig4}a.
In the presence of impurity atoms localized at one of the edges, the metallic character, with almost constant
transmission, is preserved for energies very close to $E_{F}$. For both kinds of impurities, transmission shows a narrow dip below the Fermi level for
spin-up electrons, and similar dip above the Fermi level for spin-down electrons, see Figs~\ref{fig4}a,b.
Additional and more pronounced dips appear for energies close
to $\pm$ 0.5 eV. These dips appear for both spin orientations, but are slightly separated in
energy. It is also interesting to note that in the case of P impurities a wider dip
corresponds to holes ($E - E_{F} <0$), whereas
for Al impurities the well defined and wide dip appears for electrons ($E - E_{F} >0$).

When the
impurities are shifted towards the center of the nanoribbon, some important modifications in the transmission function appear. Interesting behavior can be noticed for the PM impurity configuration (Figs~\ref{fig4}c,d). One spin channel (spin-down for Al and
spin-up for P) remains then conductive (metallic) in the close vicinity of the Fermi level,
whereas the second channel becomes semiconducting due to a wide dip in the transmission at the Fermi level.  One may expect
that the dip appears due to pronounced Fano antiresonance. This behavior significantly changes
in the PC configuration, with the impurity atoms localized in the center of the nanoribbon (Figs.~\ref{fig4}e,f). Both spin channels are then
conductive in the close vicinity of the Fermi level and metallic character of the system is recovered. However,
pronounced dips in the transmission spectrum appear for electrons (P impurities) and holes (Al impurities). It is worth to note that the Fano
antiresonances appear for energies well above or below the Fermi level. In particular, for energies close to 0.5 eV positive
interference effects in nanoribbons with P defects lead to a very well defined peak, which is followed by
a wide gap resulting from destructive interference, see the inset to Fig.~\ref{fig4}f.
Since the transmission is strongly spin dependent, the resonances for spin-up and spin-down channels are well
separated. For Al impurities, narrow Fano dips appear for both
spin channels, but now the effect can be observed for energies below the Fermi level and corresponds to the
holes. Thus, changing the impurity type one can observe quantum interference effects for electrons or holes.

Transport properties of nanoribbons with (and without) defects are fully determined by the transmission function. Thus,
having found  $T_{\sigma}(E)$ one can determine not only electric conductance, but also thermoelectric parameters. This will be presented in the following sections, and we begin with the low energy (ground) state.

\section{Electric and thermoelectric properties of ${\rm z}$S${\rm i}$NR${\rm s}$ in the low energy  state}

In the low-energy state of pristine zSiNRs, the two spin channels are equivalent. However, since  a nonzero net magnetization generally  appears in the presence of impurities, these two channels are then no longer equivalent. When the spin
channels are mixed in the nanoribbon on a distance comparable to the system length, no spin thermopower can be observed and  only conventional
thermoelectric phenomena can occur. We will consider first this limit.

In the linear response regime, the electric $I$ and heat  $I_{Q}$ currents
flowing through the system from left to right when the difference in electrical potential and temperature of the left
and right electrodes is $\Delta$V and $\Delta$T, respectively, can be written in the matrix form as~\cite{22}
\beq
\left(
\begin{array}{c}
I   \\
I_{Q} \\
\end{array}
\right)
=
\left(
\begin{array}{cc}
e^2 L_{0} & \frac{e}{T}L_{1} \\
 eL_{1} &  \frac{1}{T}L_{2} \\
\end{array}
\right)
\left(
\begin{array}{c}
\Delta V   \\
\Delta T \\
\end{array}
\right),
\eeq
where $e$ is the electron charge, while
$L_{n} = \sum_{\sigma}L_{n \sigma}$, with $L_{n \sigma} = -\frac{1}{h} \int dE\,T_{\sigma}(E)\, (E-\mu)^{n} \frac{\partial f}{\partial E} $
for $n=0,1,2$. Here, $T_{\sigma}$(E) is the spin-dependent transmission function for the system and
$f(E-\mu)$ is the Fermi-Dirac distribution function corresponding to the chemical potential $\mu$
and temperature $T$.

Basic transport coefficients can be expressed in terms of $L_{n}$.
The electrical conductance $G$ is given by the formula
$G=e^{2}L_{0}$, whereas the electronic contribution to the thermal conductance, $\kappa_{e}$, is equal to
\begin{equation}
\kappa_{e}=\frac{1}{T} \left(L_{2} - \frac{L_{1}^{2}}{L_{0}}\right).
\end{equation}
In turn, the thermopower, $S=-\Delta V /\Delta T$, is expressed by the formula
\begin{equation}
S=-\frac{L_{1}}{|e|TL_{0}}.
\end{equation}
In the linear response regime considered in this paper, transport properties are determined by electronic
states near the Fermi level. In reality the chemical potential in nanoribbons (measured from the
Fermi energy $E_F$) can be easily varied  with an external gate voltage.
This technique offers a unique possibility to realize various positions of the Fermi level in a single
sample. Alternatively, the chemical potential can be moved down or up by p-type or n-type
doping, which results in $\mu <$0 and $\mu >0$, respectively. We assume that this doping does not influence transmission functions calculated above,  and the donors/acceptors are different form the substituted P and Al atoms. Significant changes in the chemical potential can
be caused by a substrate, too.  

\begin{figure*}[ht]
 \begin{center}
    \begin{tabular}{cc}
      \resizebox{70mm}{!}{\includegraphics[angle=270]{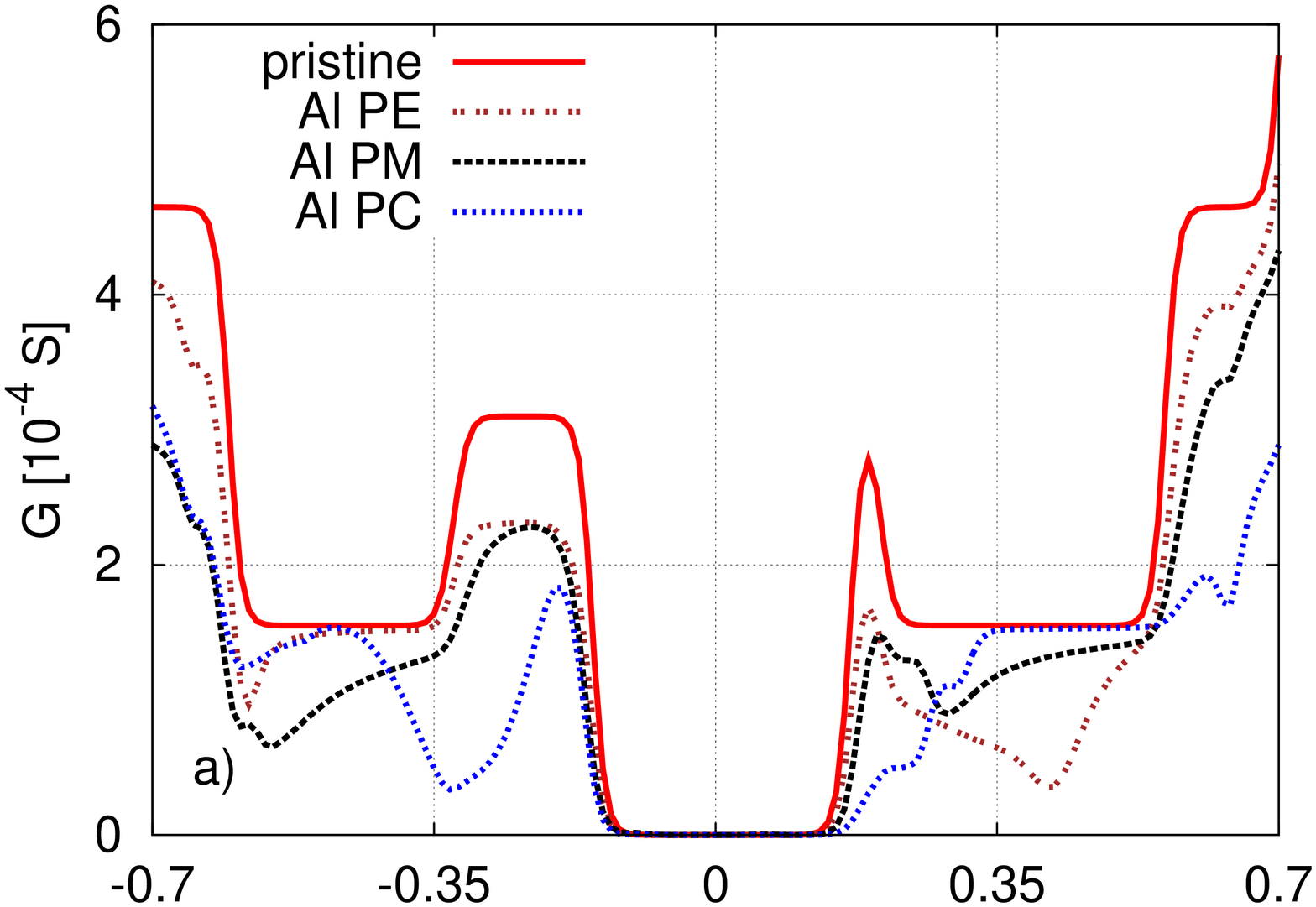}} &
      \resizebox{70mm}{!}{\includegraphics[angle=270]{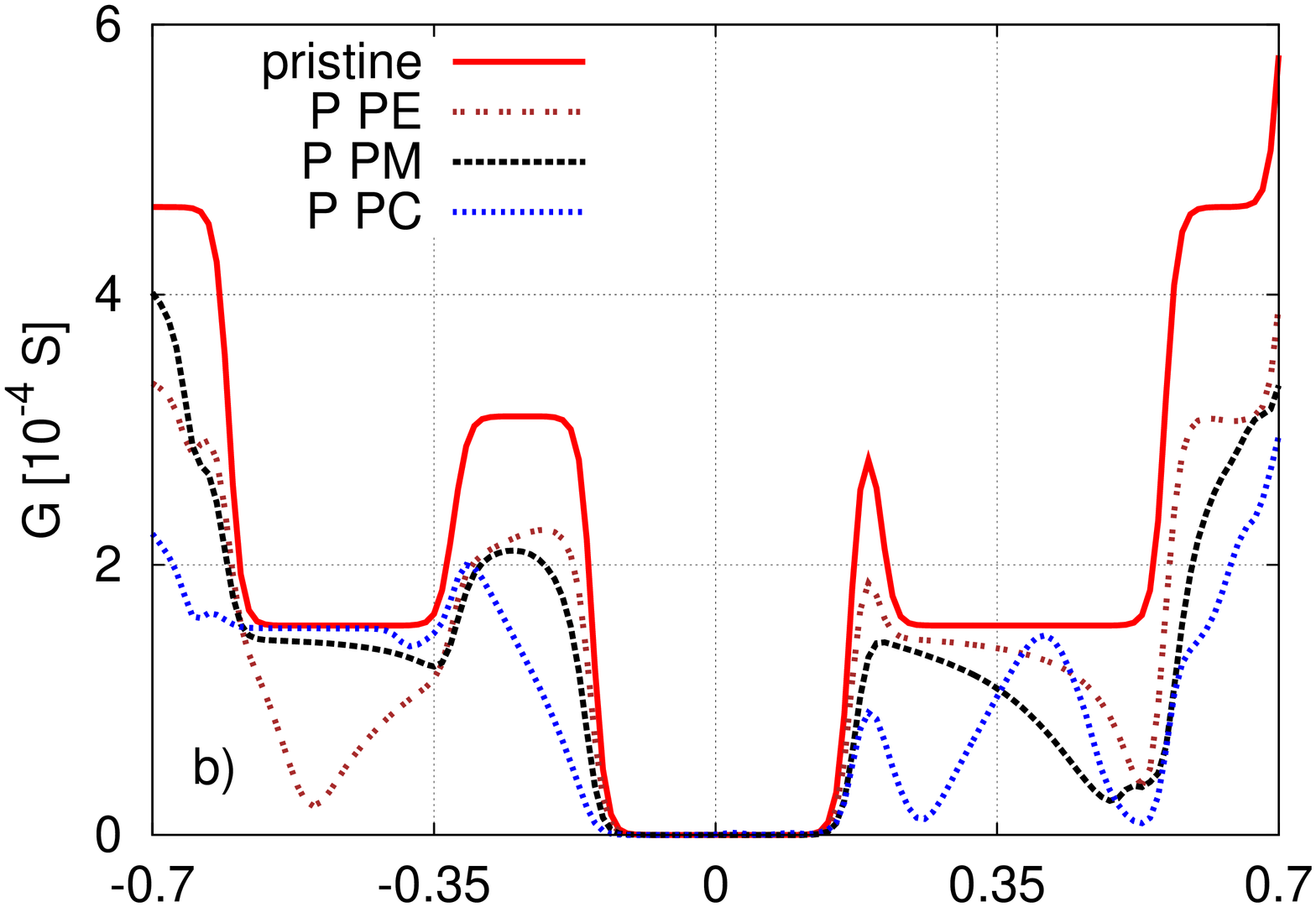}} \\
      \resizebox{70mm}{!}{\includegraphics[angle=270]{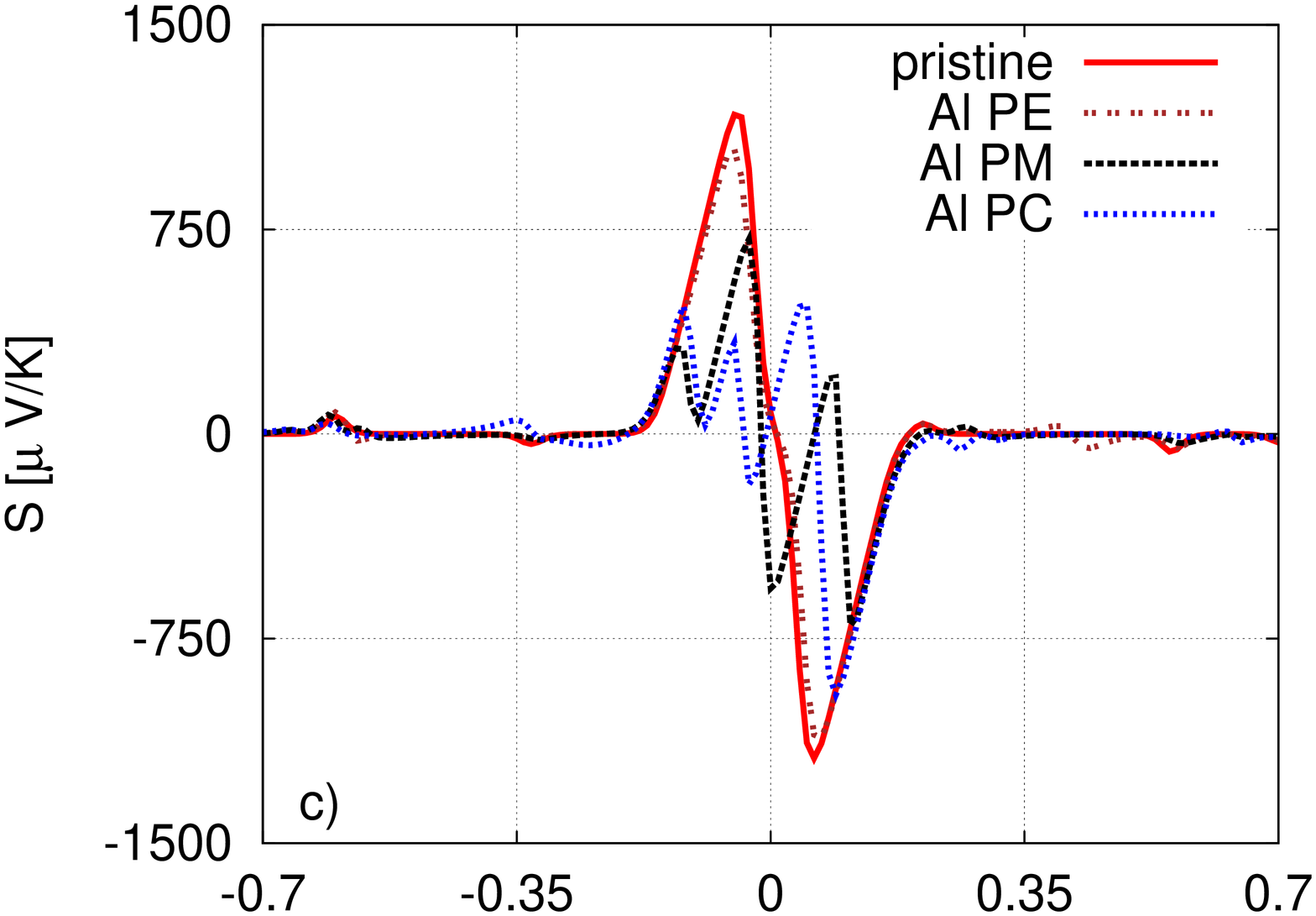}} &
      \resizebox{70mm}{!}{\includegraphics[angle=270]{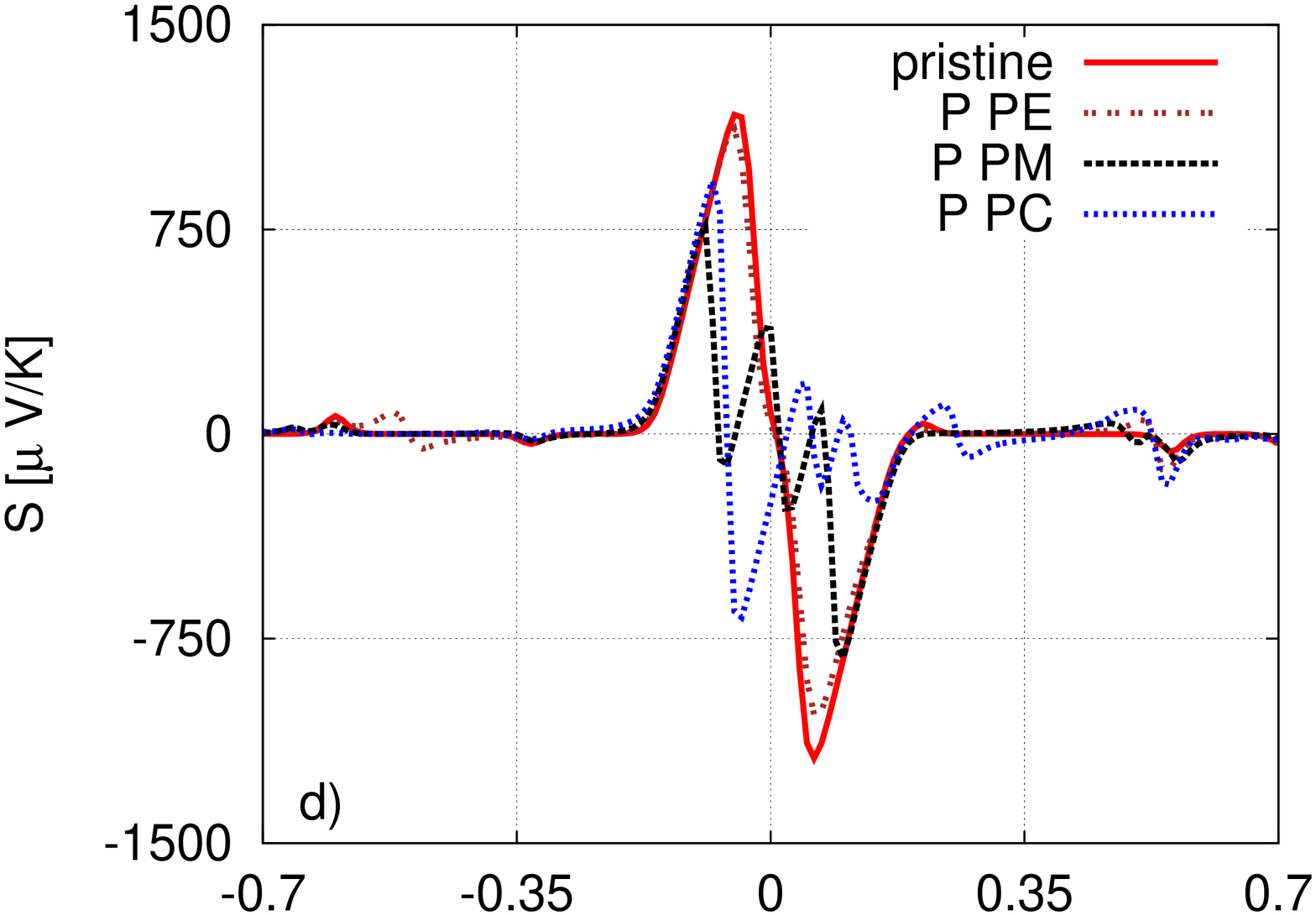}} \\
      \resizebox{70mm}{!}{\includegraphics[angle=270]{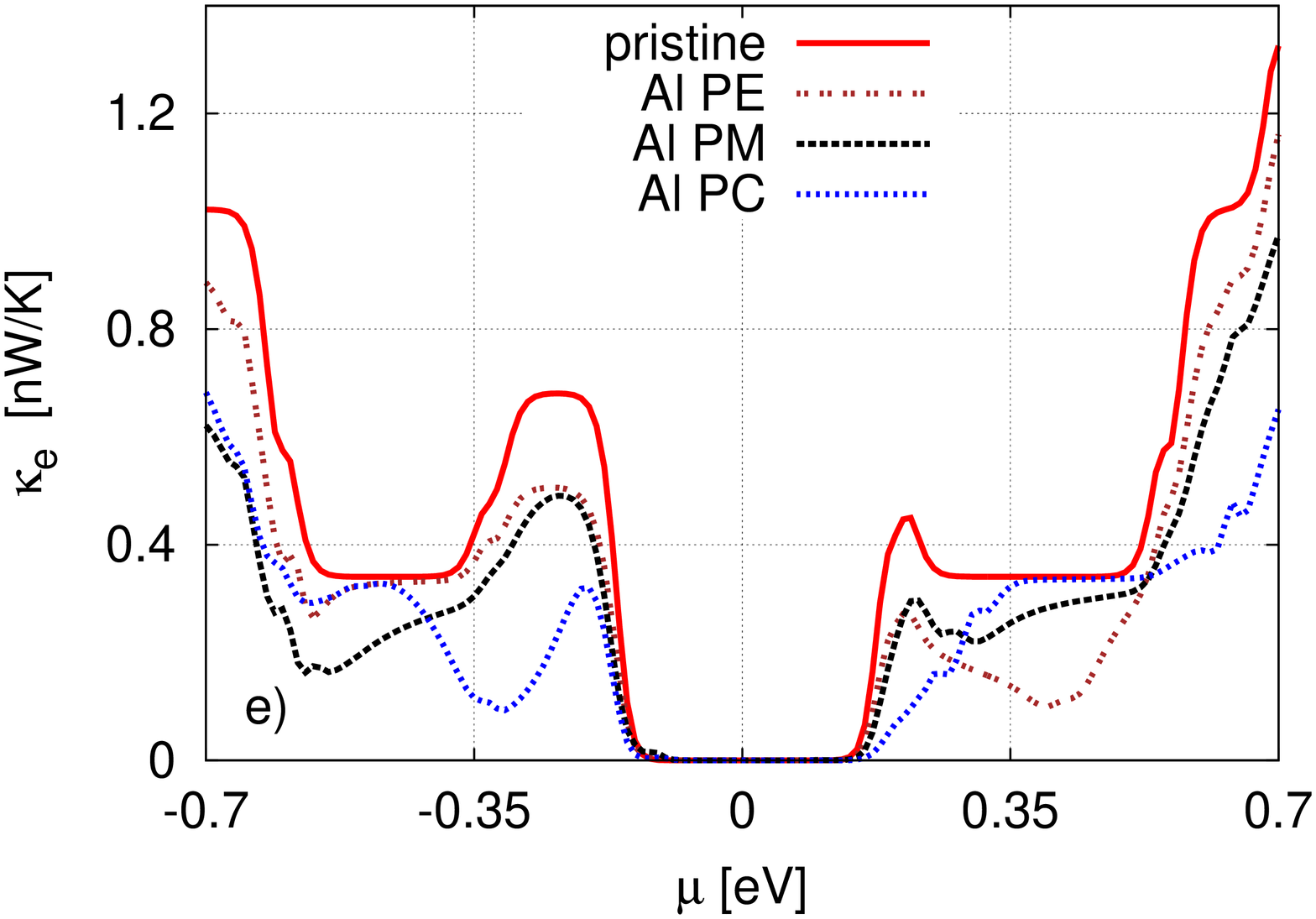}} &
      \resizebox{70mm}{!}{\includegraphics[angle=270]{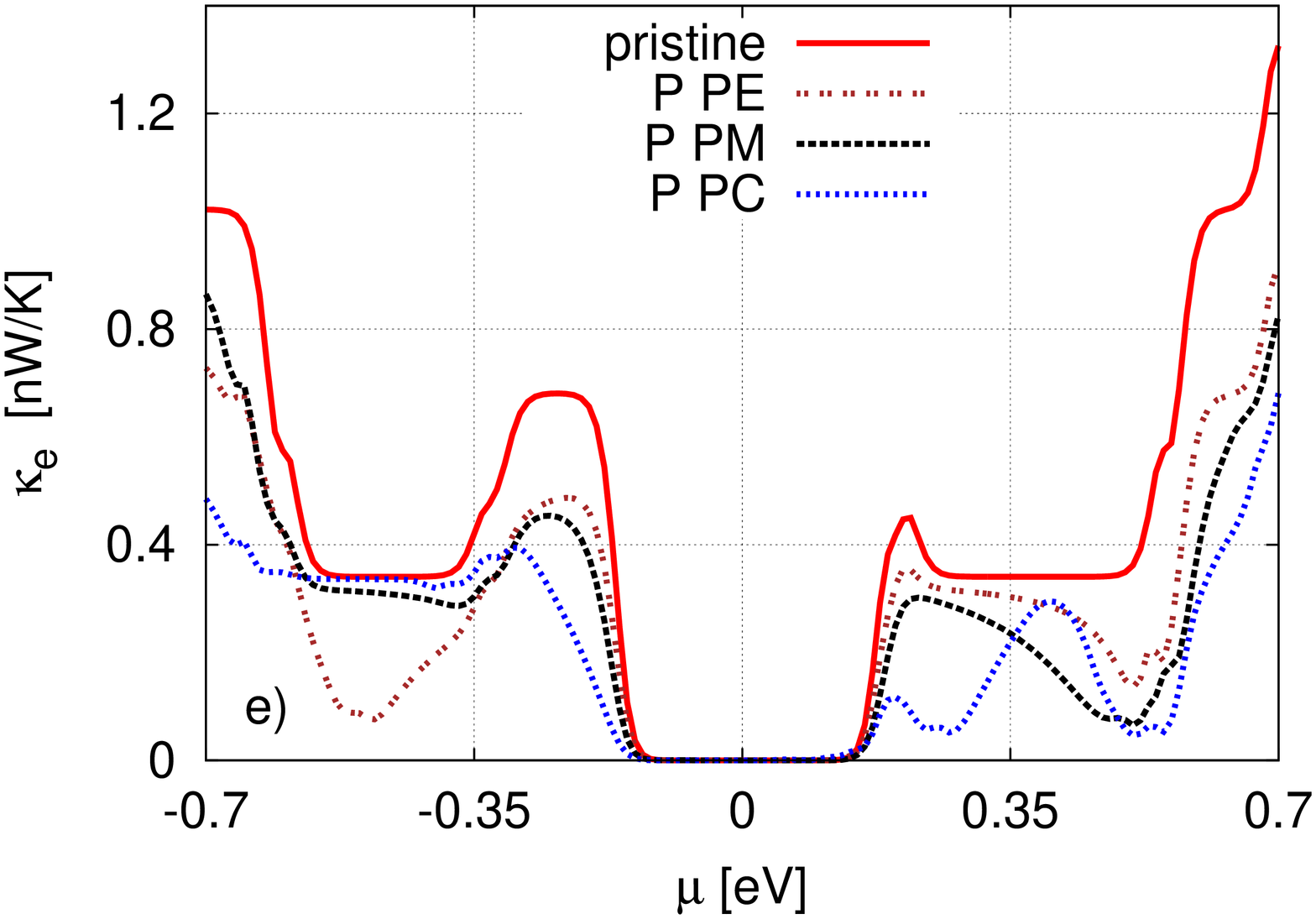}} \\
    \end{tabular}
    \caption{(Color online) Conductance (a,b), thermopower $S$ (c,d) and electronic term in the thermal conductance $\kappa_{e}$ (e,f) as a function of the chemical potential $\mu$ for zSiNRs
              in the low-energy state with Al (left panel) and P (right panel) impurities, calculated for $T=90$ K,  N=6, and the three considered impurity configurations.}
    \label{fig5}
  \end{center}
\end{figure*}

Using the transmission functions through the system (central part of the nanoribbon) determined in the previous section for the low-energy state of pristine and doped zSiNRs, one can calculate the electrical conductance $G$ and electronic term in the thermal conductance, $\kappa_{e}$, as well as the thermopower $S$. The results are presented in Fig.~\ref{fig5} as a function of the chemical potential $\mu$ for Al and P impurities, and for all the three impurity configurations (PE, PM and PC ones).
For comparison we also show there the results for pristine nanoribbons. For the temperature $T=90$ K assumed in Fig.~\ref{fig5}, the energy gap in the vicinity of $\mu=0$ is well resolved  in
the electric and in the thermal conductance. For pristine zSiNRs,  both $G$ and $\kappa_{e}$
rapidly increase  near the gap edges reaching  maximum, and then become reduced with a further increase in $|\mu |$. For higher values of $|\mu|$, the conductances
increase again. In the presence of impurity atoms, both $G$ and $\kappa_{e}$  are substantially reduced
in the whole region of the chemical potential, as shown in Figs.~\ref{fig5}a,b for $G$ and Figs.~\ref{fig5}e,f for
$\kappa_{e}$. When the impurities are localized
at the nanoribbon edge (PE configuration), the conductance shape is similar to that in pristine system,
especially near the gap, though the conductance is reduced. Moreover, a pronounced dip appears in the
region of negative $\mu$ ($\mu \approx -0.5$ eV) for P impurities and for positive $\mu$ ($\mu \approx 0.4$ eV)
for Al impurities. Much stronger modifications occur in the PC configuration, with defects localized in
the nanoribbon center. The conductances $G$ and $\kappa_{e}$ are then remarkably reduced near the gap edges,
especially for negative $\mu$ for P impurities and positive $\mu$ for Al impurities.

\begin{figure*}[ht]
 \begin{center}
    \begin{tabular}{cc}
      \resizebox{70mm}{!}{\includegraphics[angle=270]{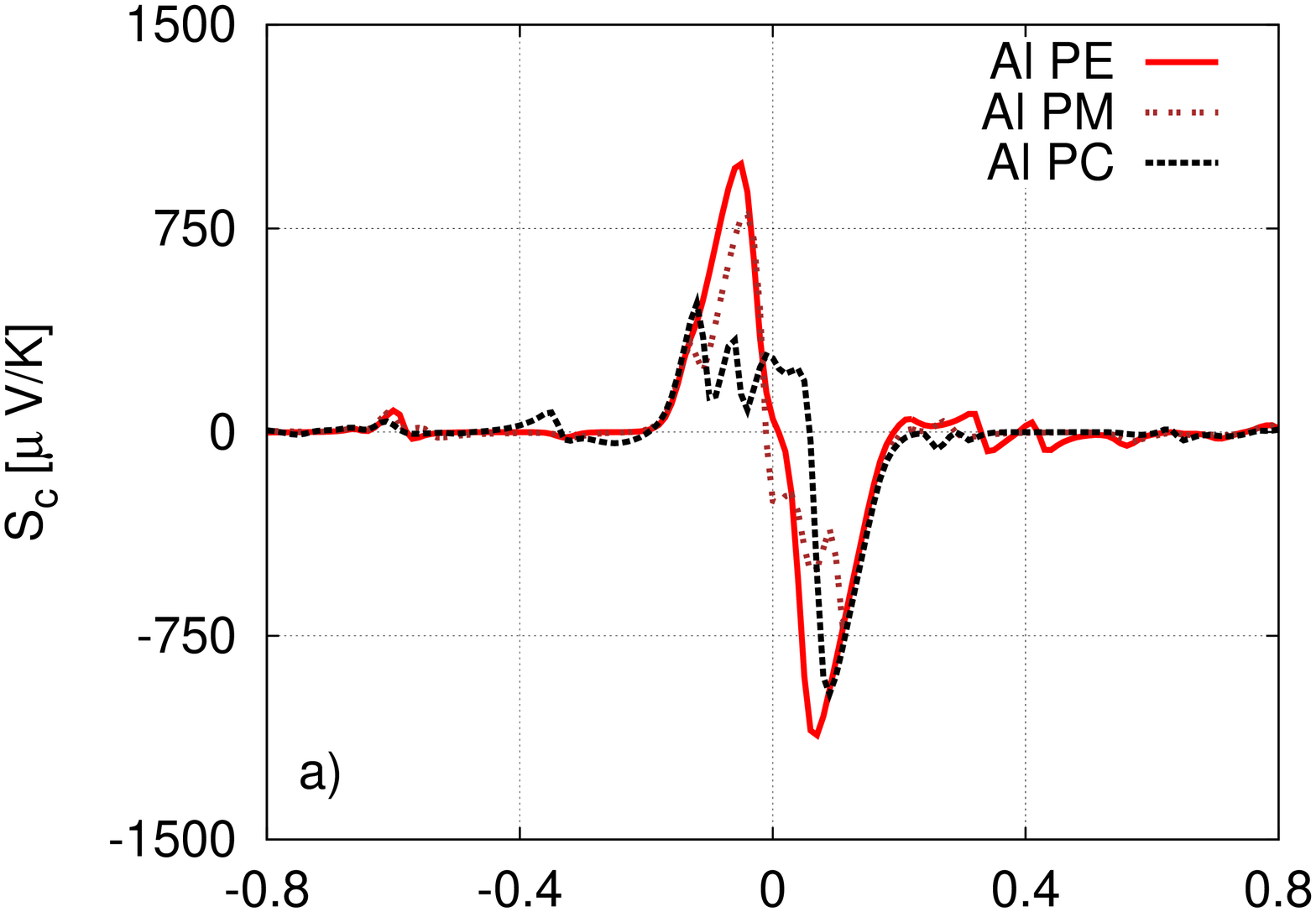}} &
      \resizebox{70mm}{!}{\includegraphics[angle=270]{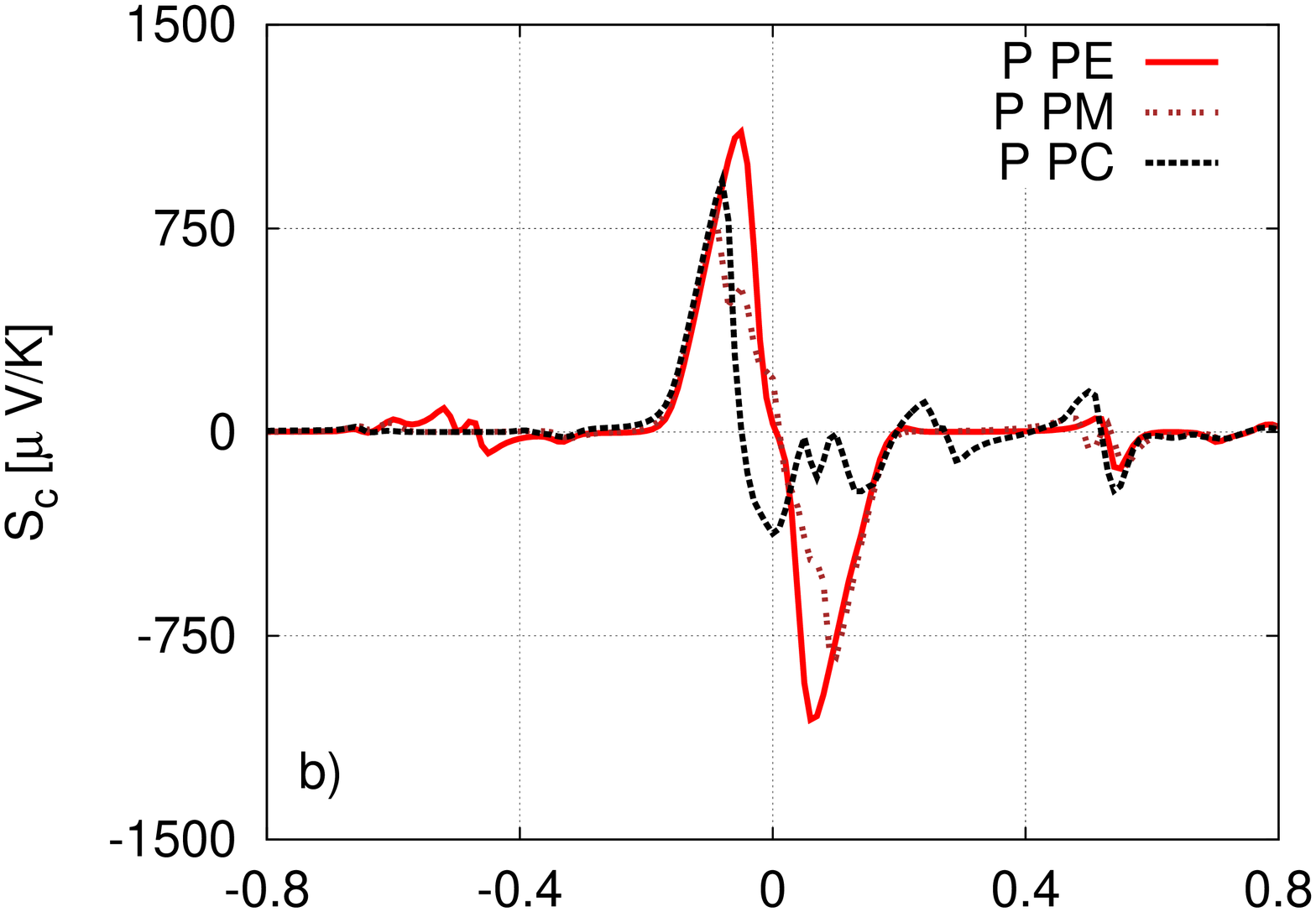}} \\
      \resizebox{70mm}{!}{\includegraphics[angle=270]{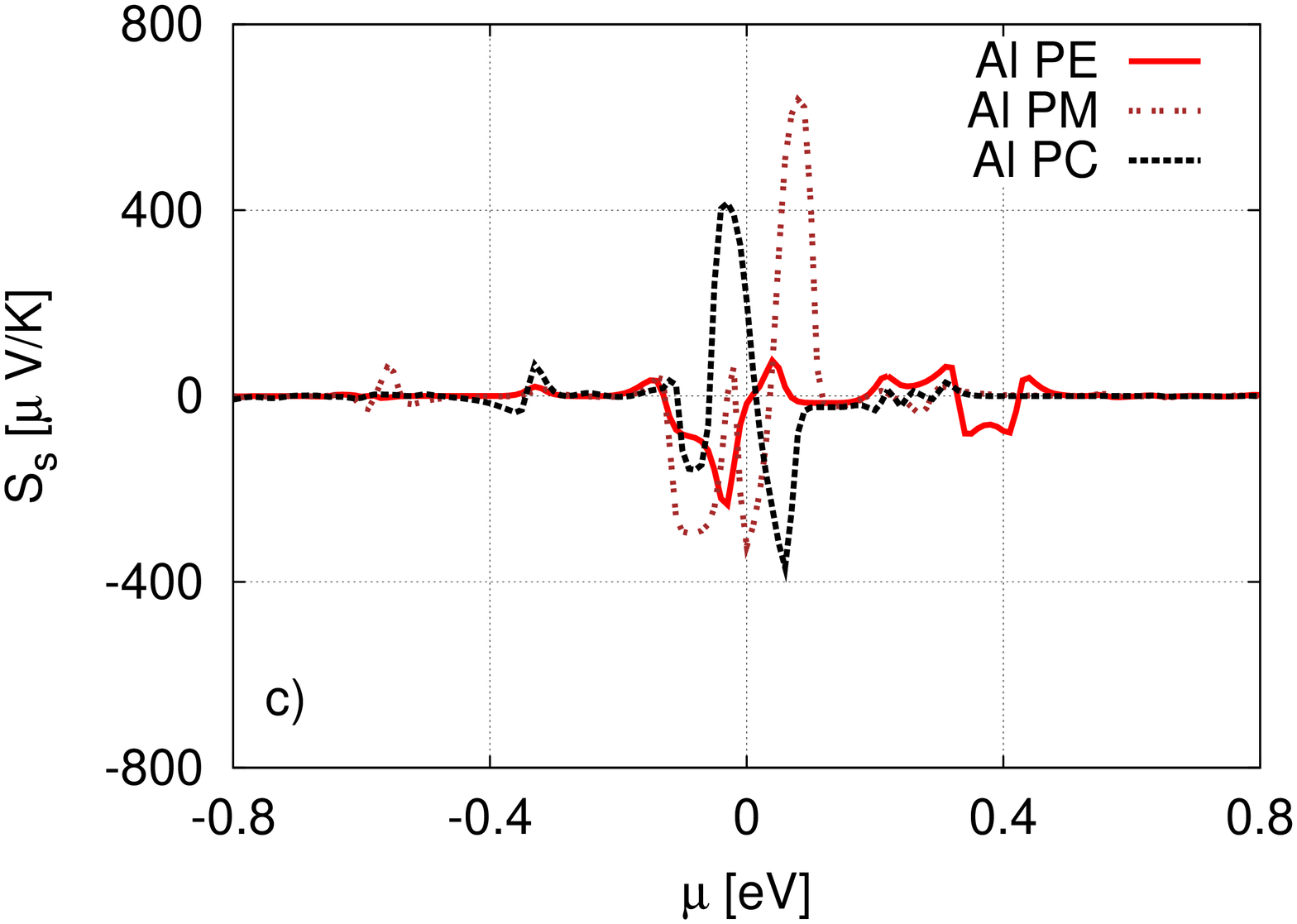}} &
      \resizebox{70mm}{!}{\includegraphics[angle=270]{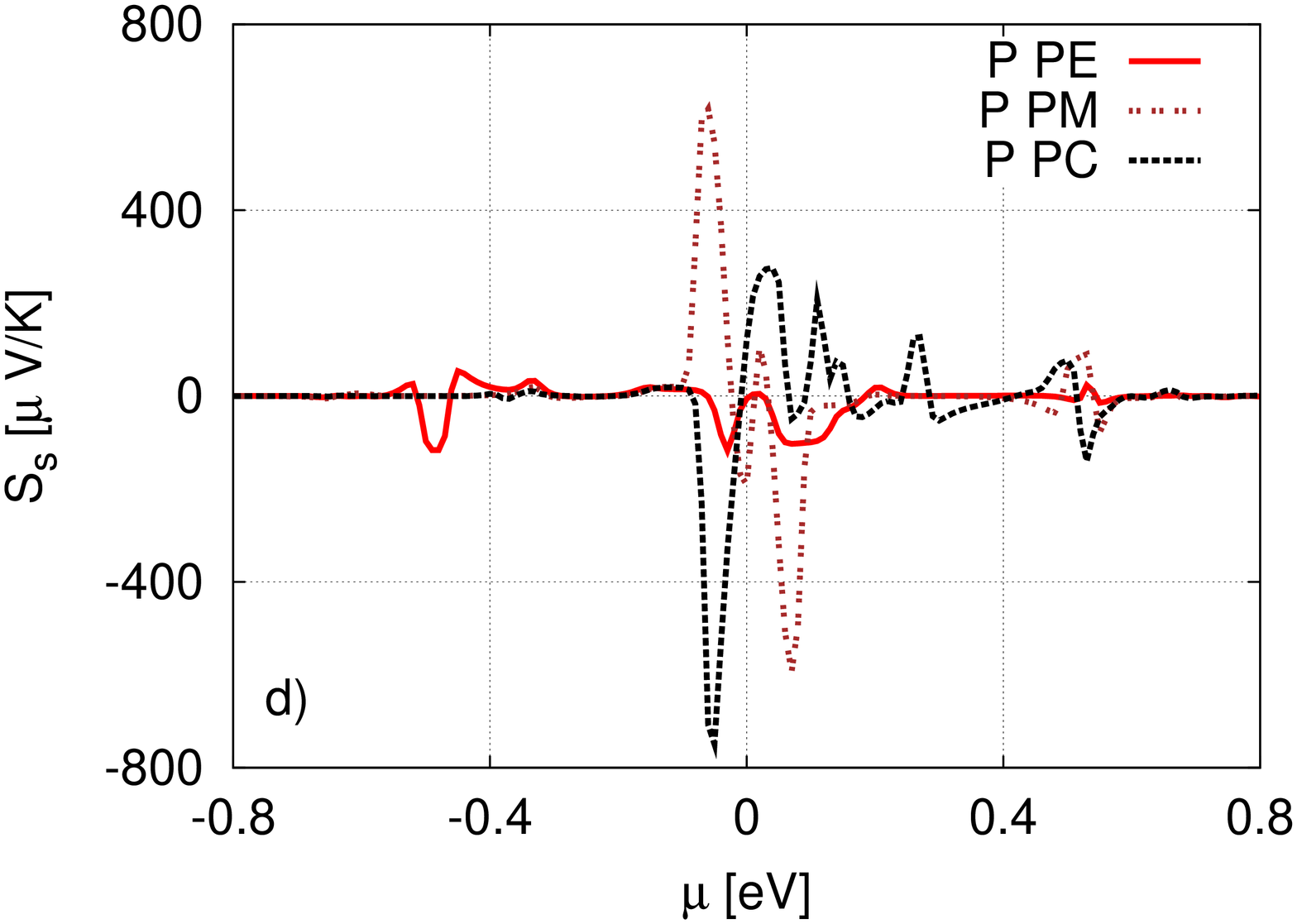}} \\
    \end{tabular}
    \caption{(Color online) Charge and spin thermopowers as a function of the chemical potential $\mu$, calculated for the low-energy  state of zSiNRs
             with Al (left panel) and P (right panel) impurities in the three (PE, PM, and PC) configurations, and for T=90K and N=6.}
    \label{figx}
  \end{center}
\end{figure*}

The thermopower $S$ of pristine nanoribbon is considerably enhanced inside the main gap in the spectrum (around $E_F$), with maxima (positive and negative) at the chemical potentials of several $kT$ from the
left and right gap edges. The maxima appear when either particle transport (for negative $\mu$) or hole transport  (for positive $\mu$) becomes suppressed, as discussed in details in Ref.~\onlinecite{22}. A large value of $|S|$, exceeding 1mV/K, results from rapid increase
in transmission near the gap edges. Similar enhancement of the thermopower also appears in the presence of
impurities in the PE configuration. However, for nanoribbons with impurities in the PC configuration, the  thermopower
is considerably reduced since transmission near the gap edges is diminished and changes more smoothly.
Moreover, a kind of damped oscillations of $S$ inside the gap can be observed, which follow from states localized in the gap. It is interesting to note that
the global maximum of the thermopower in the presence of P defects in the PC configuration is positive and appears  near the left edge of the gap (negative $\mu$), whereas such a global maximum in the presence of Al impurities is negative and appears near the right edge of the gap (positive $\mu$). In the presence of impurities,
the thermopower is also enhanced for higher values of $|\mu|$. This enhancement, however, is less pronounced than that for  $\mu$ inside the main gap.

Situation may change when the spin channels are ether not mixed in the nanoribbon, or they are mixed on a scale much longer than the system's length. The spin effects in thermoelectric properties become then important.
The electrical conductance $G_{\sigma}$ of the spin-$\sigma$ channel
is equal to $e^{2} L_{0\sigma}$, $G_{\sigma} = e^{2} L_{0\sigma}$, whereas the total electronic contribution to the thermal conductance,
$\kappa_{e}$, is given by the formula
\begin{equation}
\kappa_{e}=\frac{1}{T} \sum_{\sigma} \left(L_{2\sigma} - \frac{L_{1\sigma}^{2}}{L_{0\sigma}}\right).
\end{equation}
Since  the two spin channels are not mixed and spin accumulation is important, one can introduce spin-dependent
thermopower,
$S_{\sigma}=-\Delta V_{\sigma} /\Delta T = -L_{1\sigma} / |e|TL_{0\sigma}$, which corresponds to the spin-dependent
voltage generated by a temperature gradient ~\cite{22,34}. The conventional (charge) and spin thermopowers can be then written as
$S_c = \frac{1}{2}(S_{\uparrow} + S_{\downarrow})$ and $S_{s} = \frac{1}{2}(S_{\uparrow} - S_{\downarrow})$, respectively.

We assume the transmission through the system (central part of the nanoribbon)  is the same for nanoribbons with and without spin relaxation. In Fig.~\ref{figx} we show both charge and spin thermopowers as a function of chemical potential for all the three considered localizations of the Al and P impurity atoms.
It is interesting to note, that the spin thermopower is relatively large due to a significant spin dependence of the transmission function in the presence of impurities and for the assumed distance between the impurity atoms. The
impurity-induced spin thermopower depends on the location of the impurities and
is especially large in the PC and PM configurations. When the impurities are located at one of the edges, the spin thermopower is remarkably  smaller.  Note, the conventional (charge) thermopower is also modified as follows from comparison of the results shown in Fig.~\ref{figx} with
the corresponding ones in Fig.~\ref{fig5}. This modification appears due to impurity-induced spin dependence of the transmission function. Some modifications induced by impurities appear also in other transport coefficients, for instance in the electronic contribution to the heat conductance, which however are not presented here explicitly.

\section{Electric and thermoelectric properties of ${\rm z}$S${\rm i}$NR${\rm s}$ in the ferromagnetic state}

By applying an external magnetic field one can stabilize the ferromagnetic (FM) configuration of zSiNRs. This configuration can be also stabilized by exchange coupling to a ferromagnetic substrate or to ferromagnetic contacts. Due to a significant magnetic moment in the FM state, transmission  function of a nanoribbon is  strongly spin-dependent.
\begin{figure*}[ht]
 \begin{center}
    \begin{tabular}{cc}
      \resizebox{70mm}{!}{\includegraphics[angle=270]{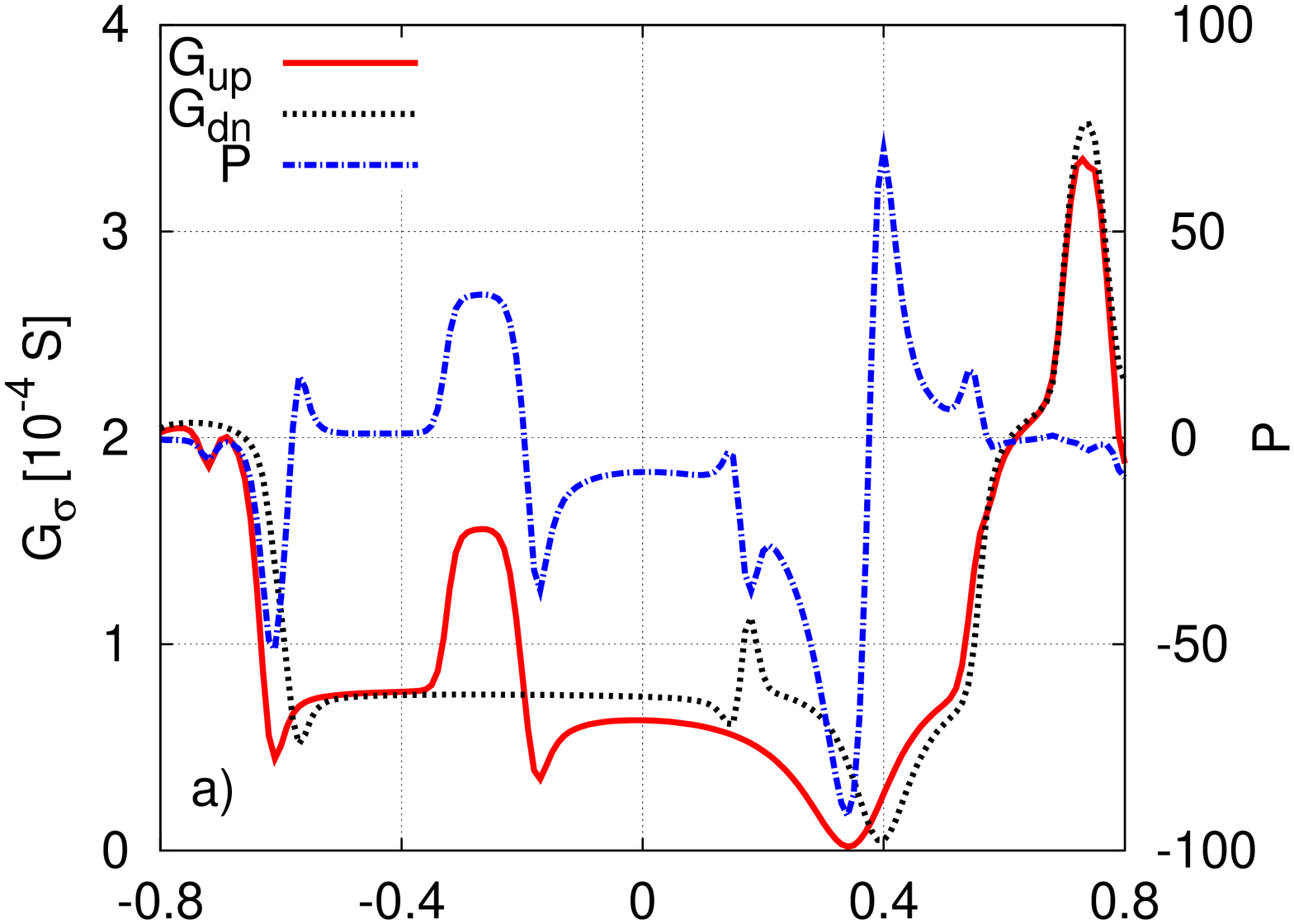}} &
      \resizebox{70mm}{!}{\includegraphics[angle=270]{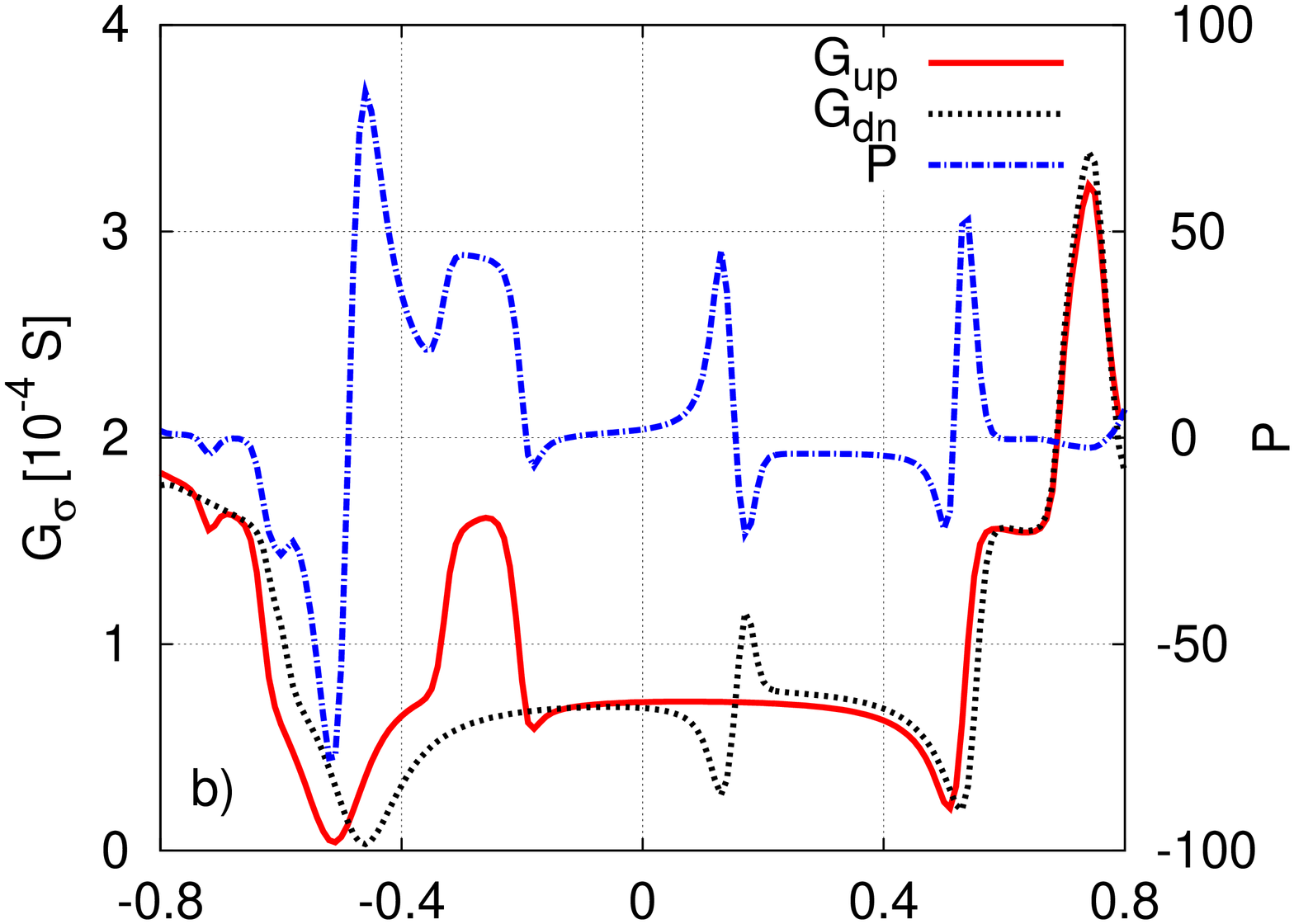}} \\
      \resizebox{70mm}{!}{\includegraphics[angle=270]{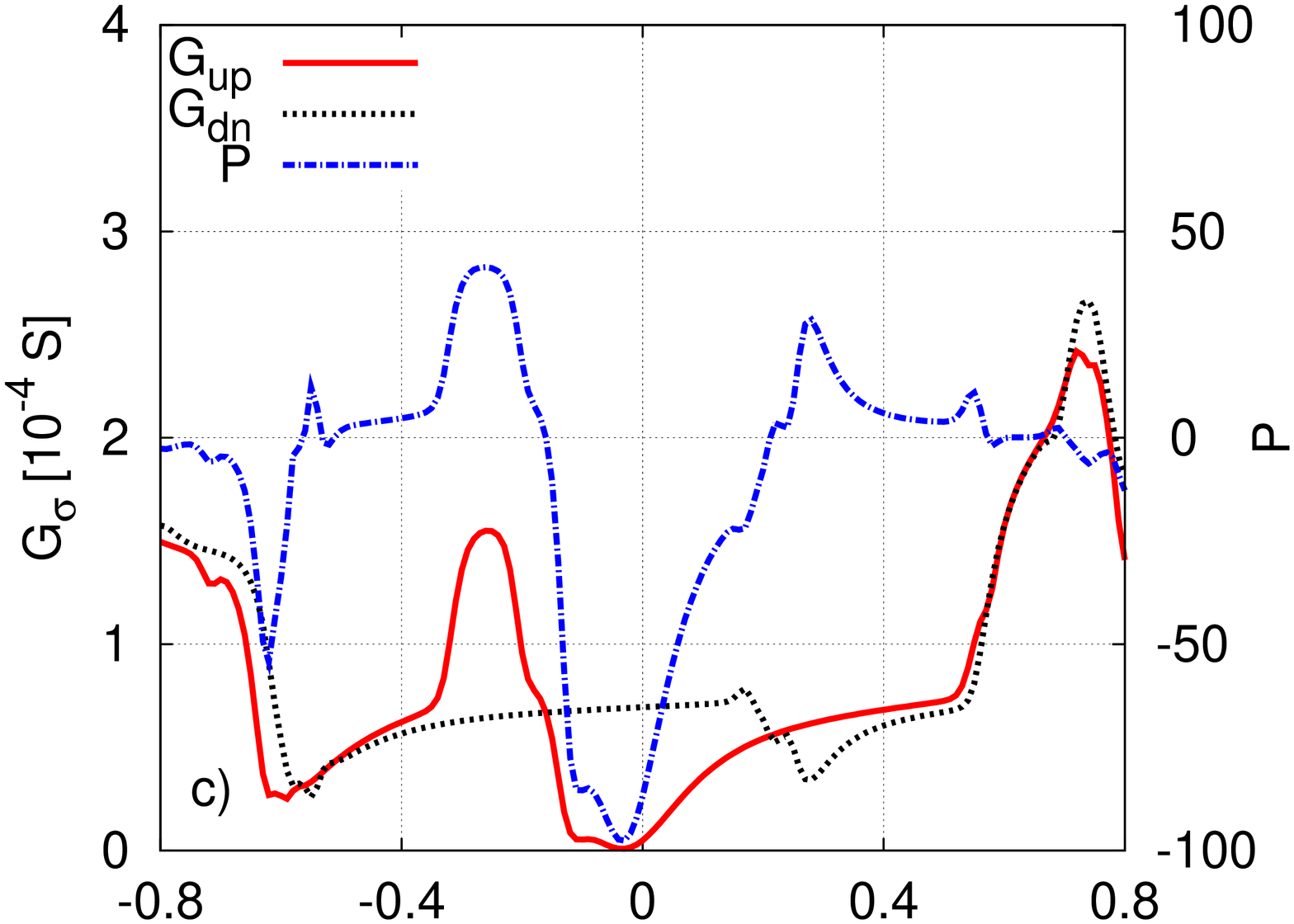}} &
      \resizebox{70mm}{!}{\includegraphics[angle=270]{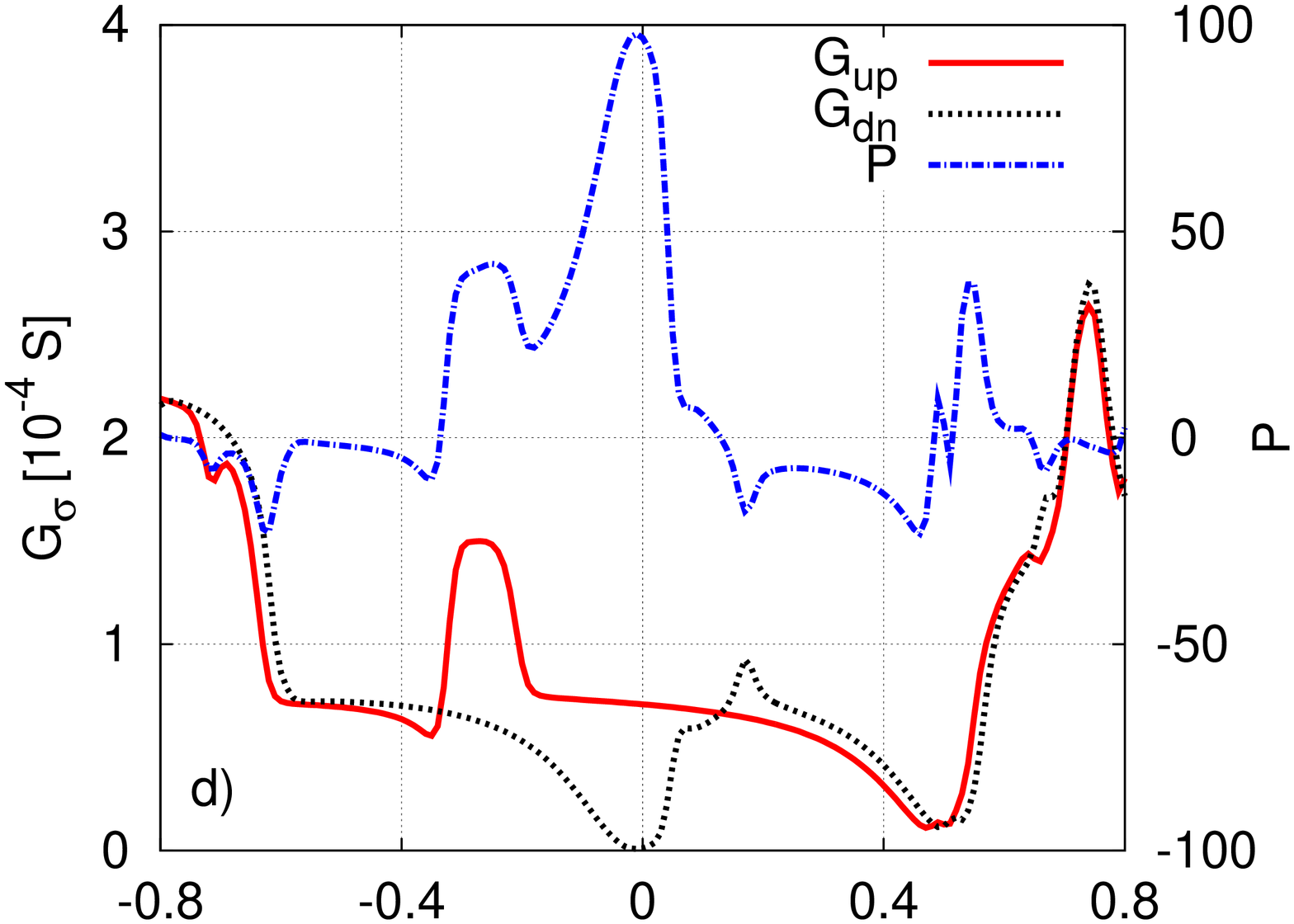}} \\
      \resizebox{70mm}{!}{\includegraphics[angle=270]{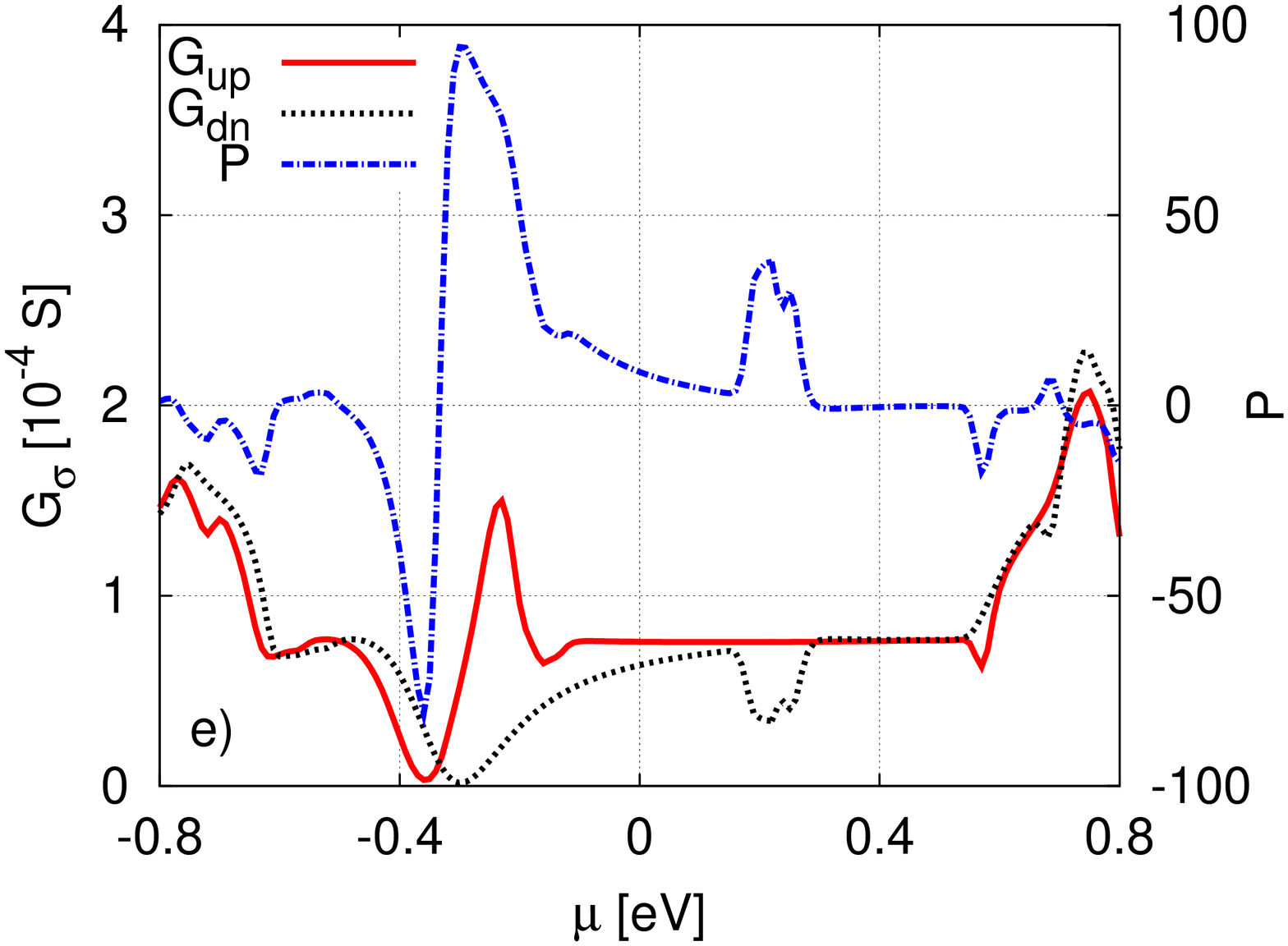}} &
      \resizebox{70mm}{!}{\includegraphics[angle=270]{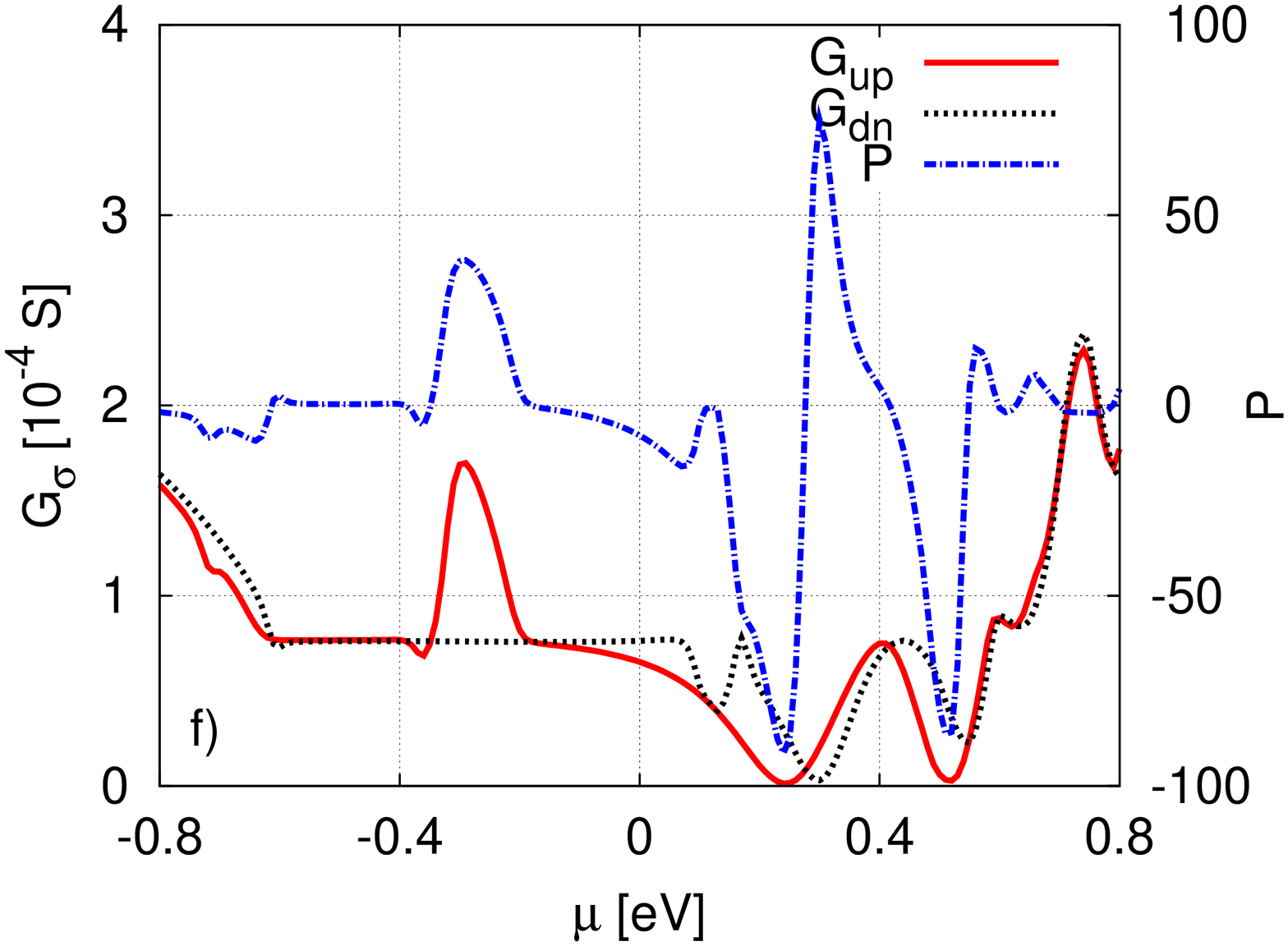}} \\
    \end{tabular}
    \caption{(Color online) Spin-dependent conductance $G_{\sigma}$ and polarization $P$ as a function of chemical potential $\mu$, calculated for the FM state of ZSiNRs
             with Al (left panel) and P (right panel) impurity atoms in the configurations PE (a,b), PM (c,d), and PC (e,f). The other parameters are $T=90$K, and $N=6$.}
    \label{fig6}
  \end{center}
\end{figure*}

\begin{figure*}[ht]
 \begin{center}
    \begin{tabular}{cc}
      \resizebox{70mm}{!}{\includegraphics[angle=270]{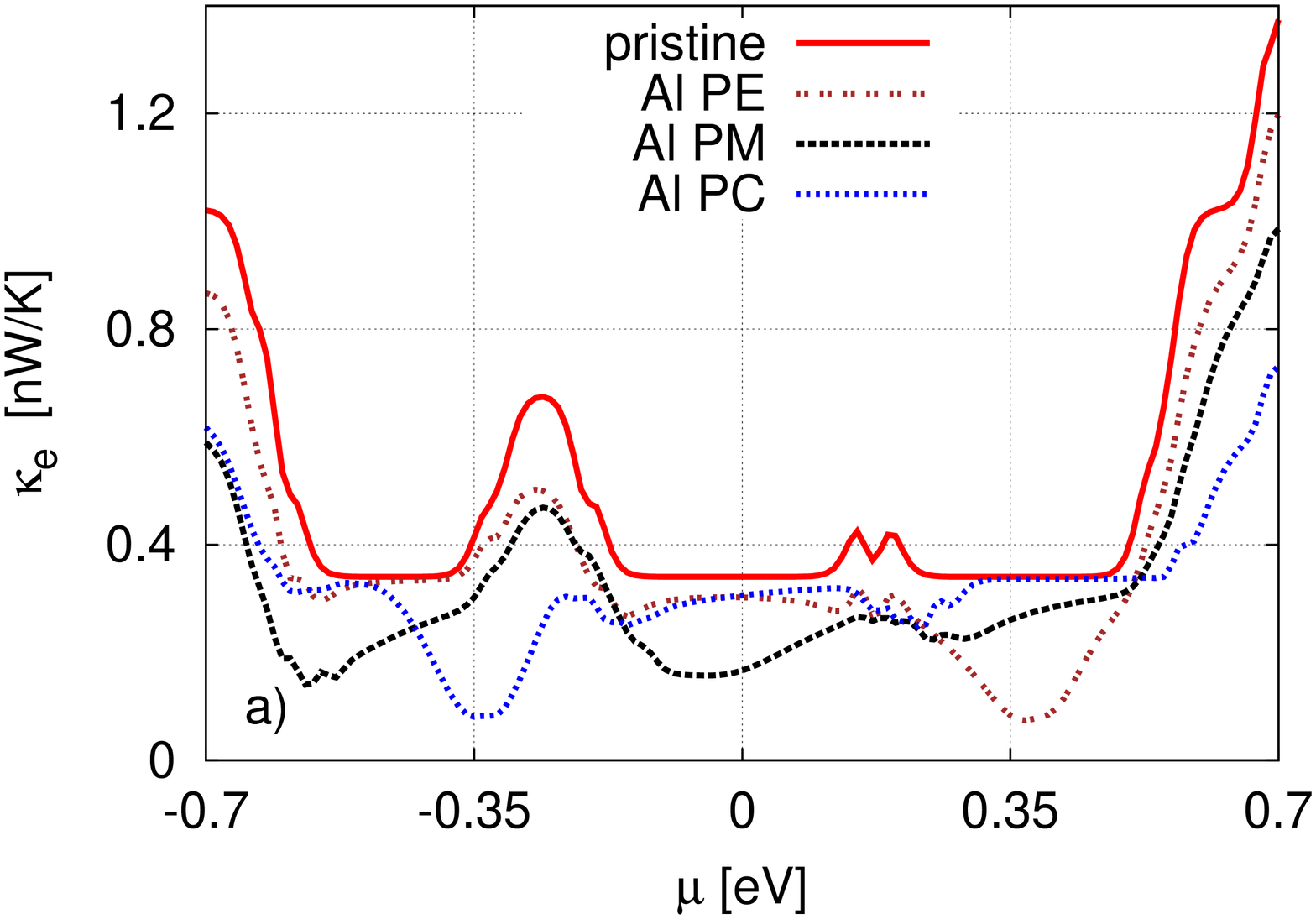}} &
      \resizebox{70mm}{!}{\includegraphics[angle=270]{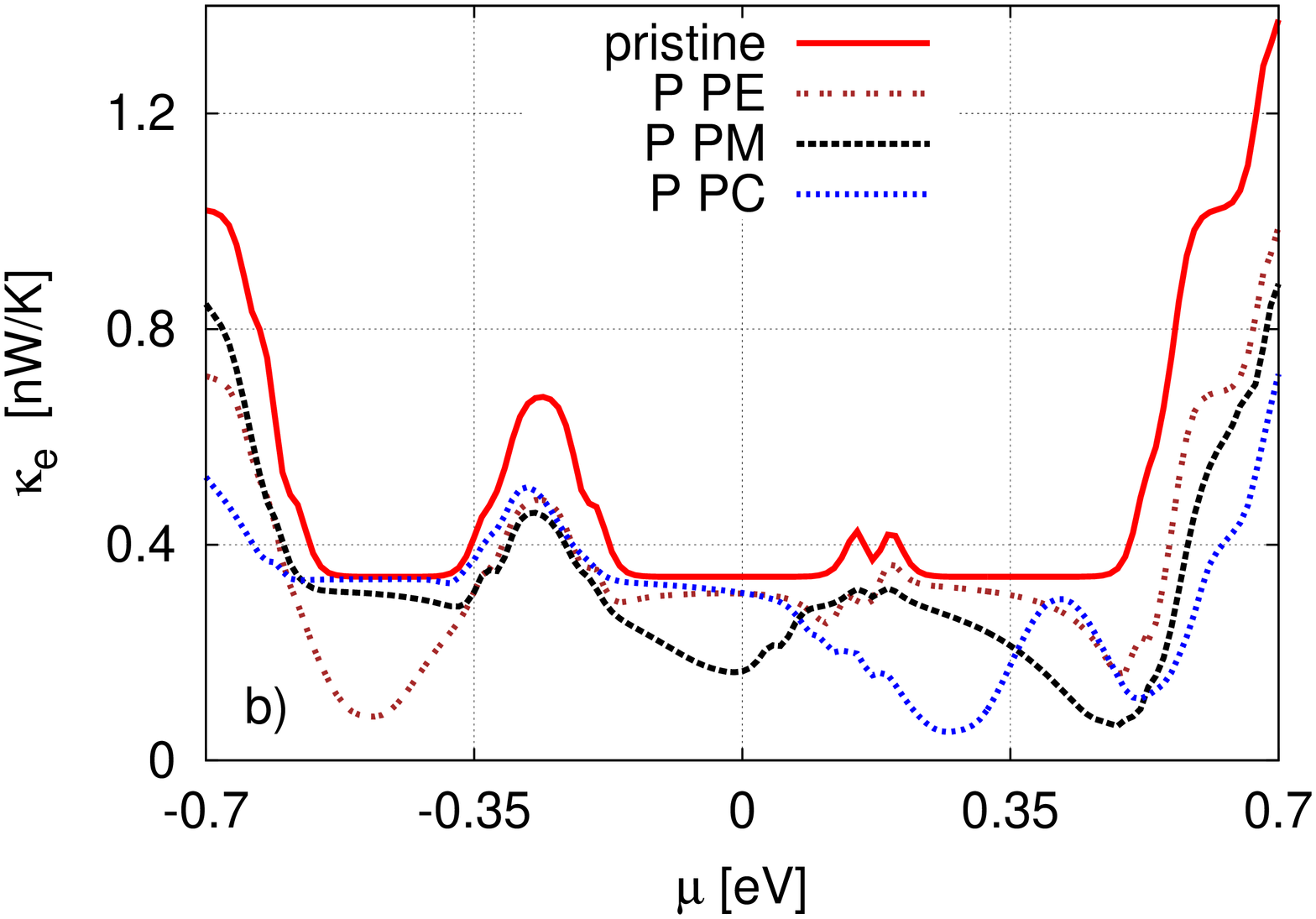}} \\
    \end{tabular}
    \caption{(Color online) Electronic term in the thermal conductance, $\kappa_{e}$, as a function of chemical potential $\mu$ in the FM state of zSiNRs for Al (a) and P (b) impurities, and for $T=90$K and  $N=6$.}
    \label{fig7}
  \end{center}
\end{figure*}

\begin{figure*}[ht]
 \begin{center}
    \begin{tabular}{cc}
      \resizebox{70mm}{!}{\includegraphics[angle=270]{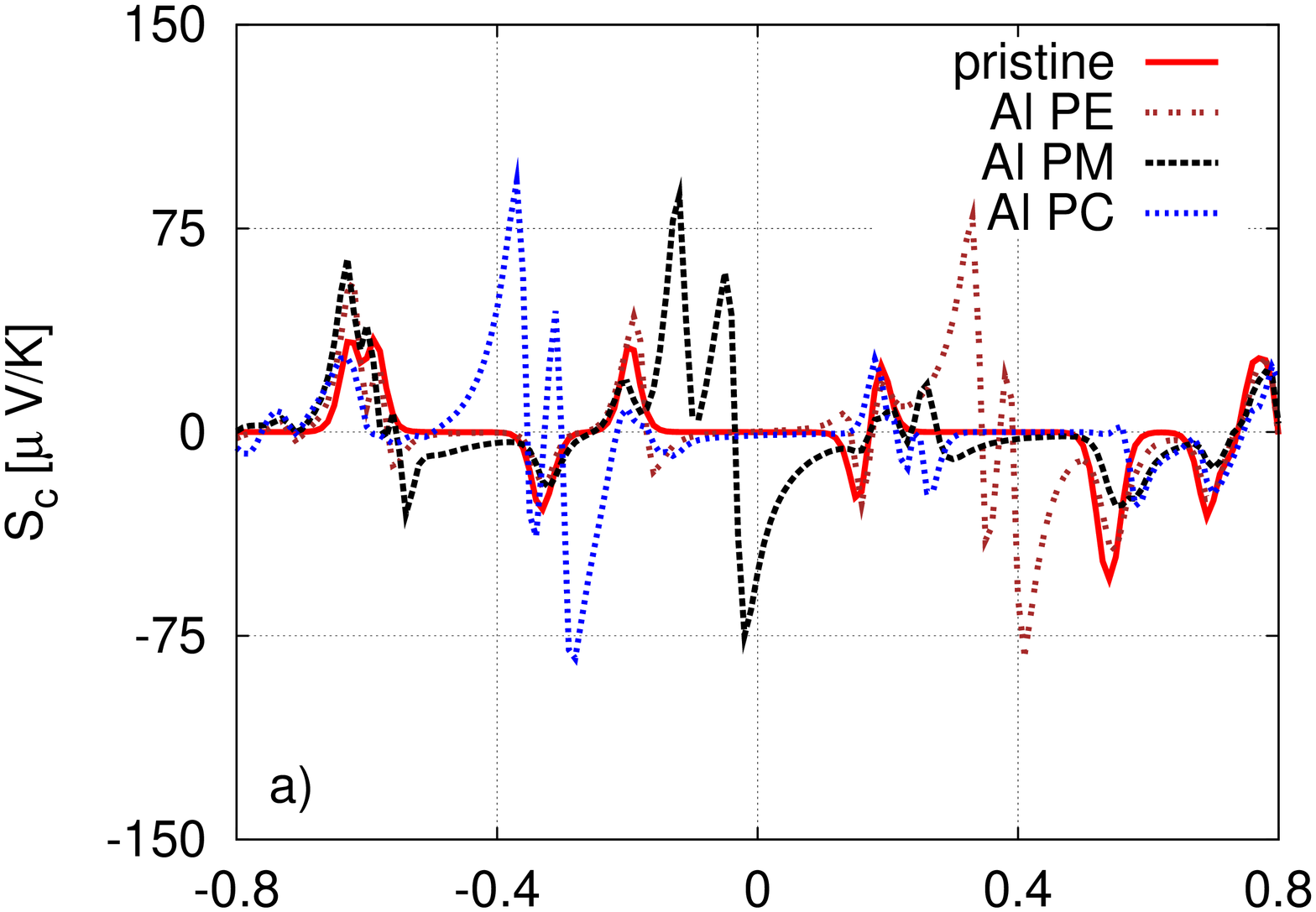}} &
      \resizebox{70mm}{!}{\includegraphics[angle=270]{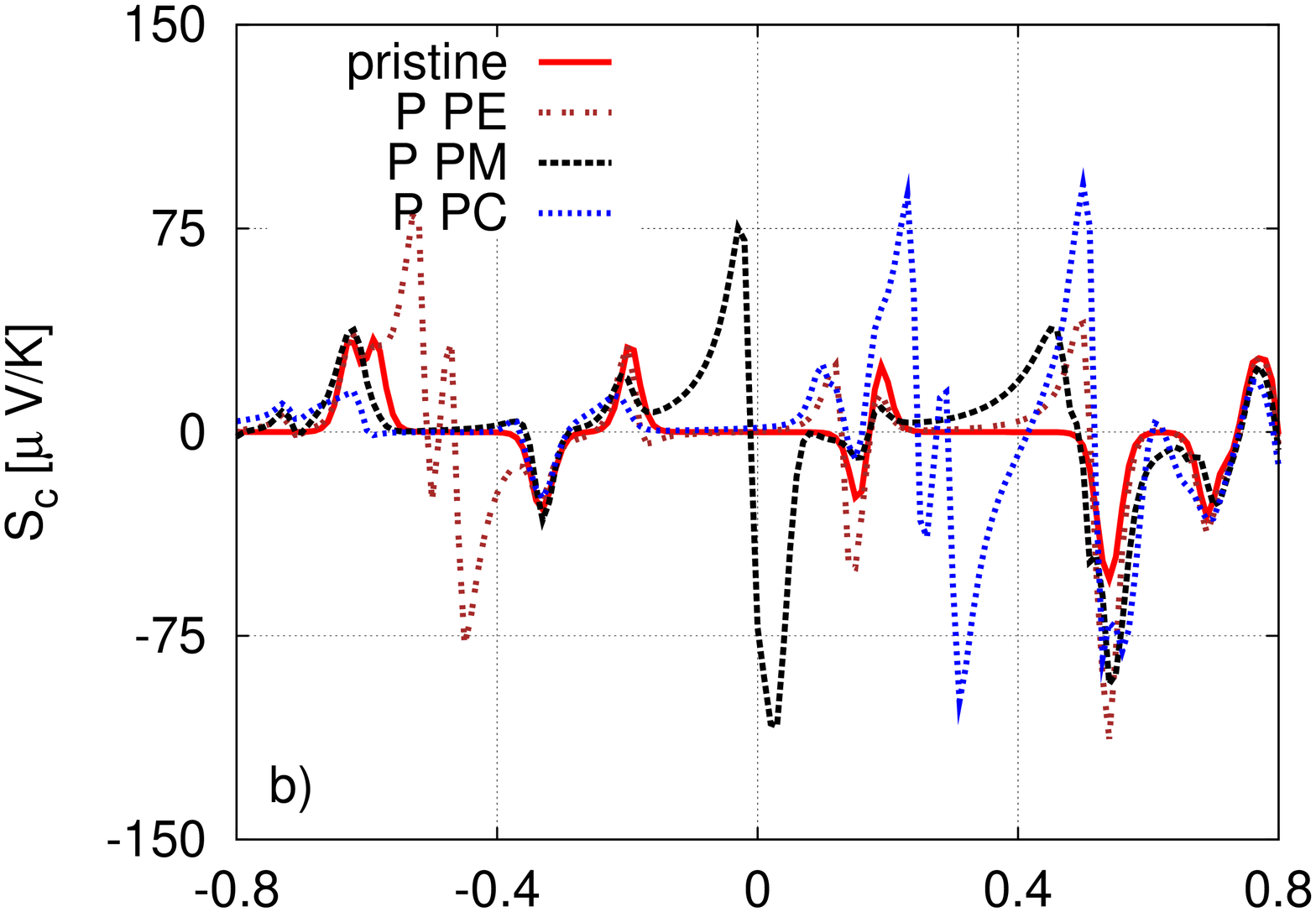}} \\
      \resizebox{70mm}{!}{\includegraphics[angle=270]{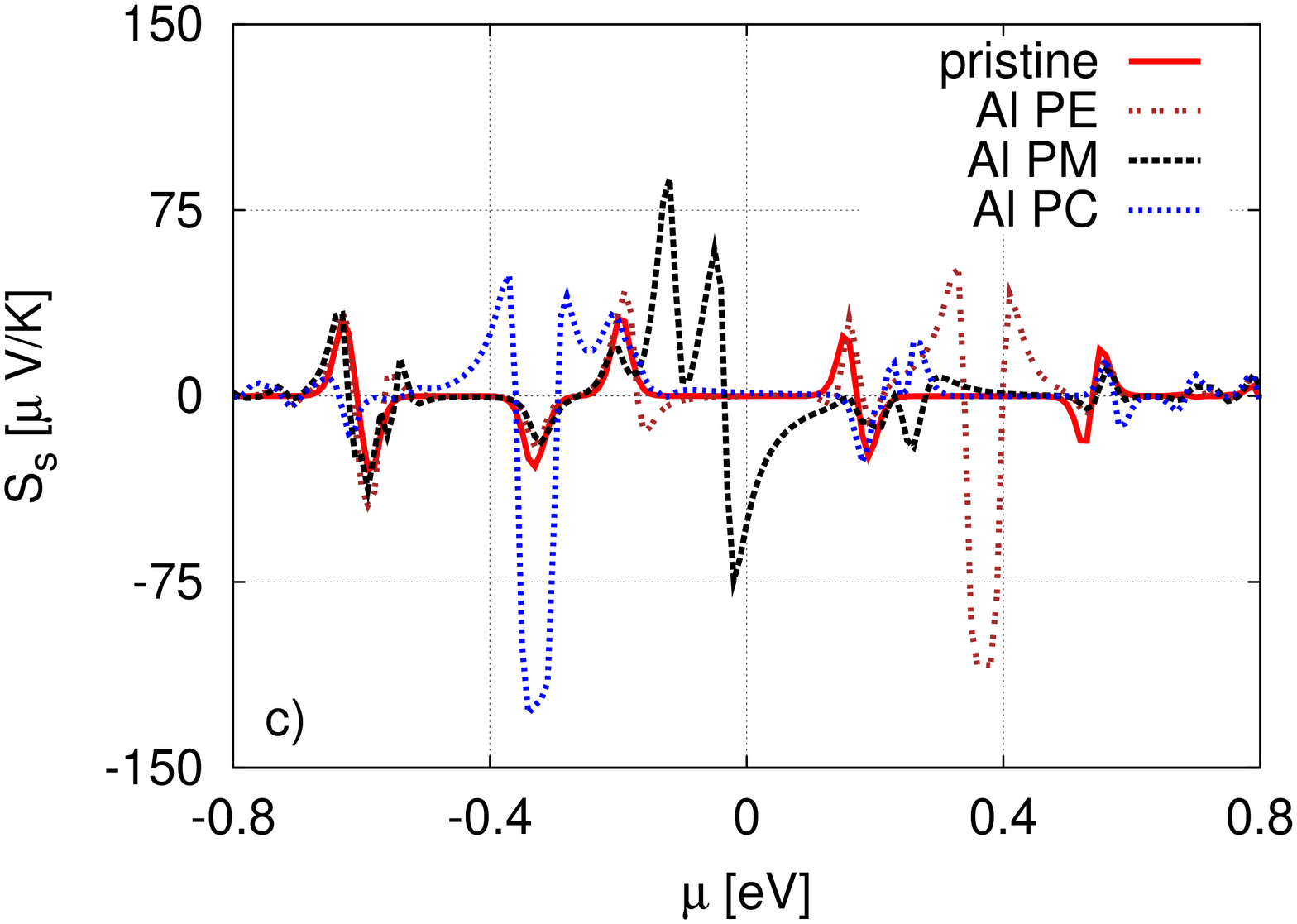}} &
      \resizebox{70mm}{!}{\includegraphics[angle=270]{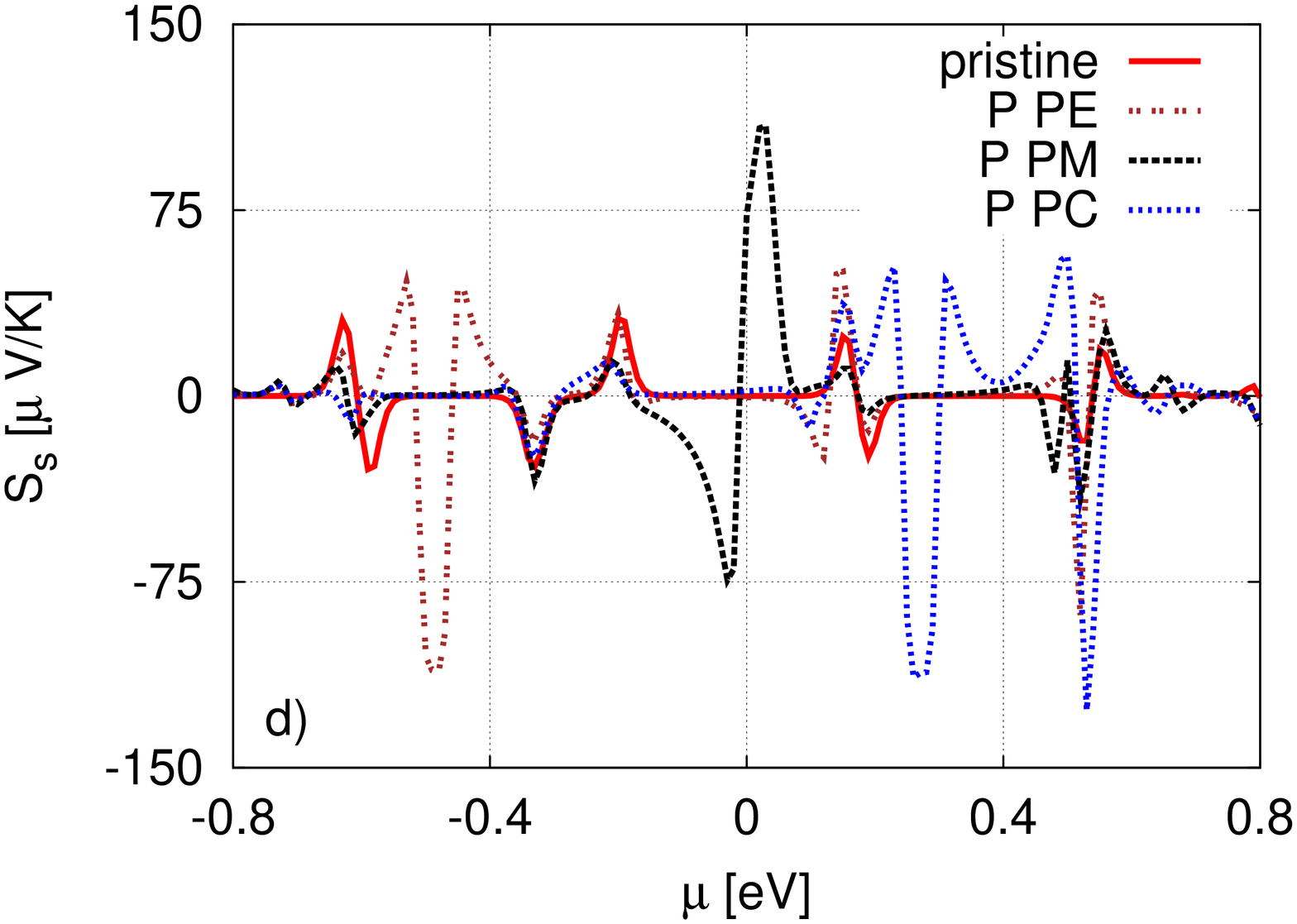}} \\
    \end{tabular}
    \caption{(Color online) Charge S$_{c}$ and spin S$_{s}$ thermopower as a function of chemical potential $\mu$ calculated for the FM state for zSiNRs with Al (a,c) and P (b,d) impurities,  and for $T=90$K and $N=6$.}
    \label{fig8}
  \end{center}
\end{figure*}

Using the spin-resolved transmission $T_{\sigma}$(E) determined in Sec. 2B for the ferromagnetic state of pristine
nanoribbons as well as of nanoribbons with impurities, we investigate now the spin-polarized transport phenomena. Spin-resolved electrical
conductance $G_{\sigma}$ as a function of chemical potential $\mu$ is presented in Fig.~\ref{fig6} for different positions
of the Al and P impurity atoms, whereas the thermal conductance $\kappa_{e}$  is depicted in Fig. ~\ref{fig7}. Pristine
ferromagnetic nanoribbons as well as those with impurities in the PE configuration show metallic character. The conductance
$G_{\sigma}$ is then constant in a close vicinity of $\mu=0$ and it is practically the same for both spin channels
for P impurities, and weakly depends on spin for Al impurities. Strong modifications appear  for higher values of $|\mu|$,
where due to a pronounced dip in the transmission, one spin channel becomes weakly conductive and a considerable
polarization $P$, defined as  $P=\frac{G_{\uparrow}-G_{\downarrow}}{G_{\uparrow}+G_{\downarrow}}\times 100\%$, can
be observed. It is worth to note that the presence of P atoms at one of the edges leads to a considerable polarization
(up to 90$\%$) in a narrow region of negative $\mu$, whereas for Al impurities  a large polarization can be
obtained for positive $\mu$. For other values of chemical potential, the spin polarization is much smaller, though
in narrow regions of $\mu$ it can achieve even 50$\%$. Interesting results are obtained for the PM impurity configuration, where
one spin channel -- spin up for Al and spin down for P impurities -- becomes nonconductive in a close vicinity of $\mu$=0,
whereas the second channel exhibits metallic character. The system behaves thus like a half-metallic ferromagnet
with practically 100$\%$ polarization. Such a behavior can be important for potential applications of doped silicene
nanoribbons in spintronic devices. Relatively high polarization can be also obtained for PC configuration, but in
a narrow region of negative $\mu$ for Al impurity and positive $\mu$ for P impurities. Moreover, the conductance
of a nanoribbon with central P impurities exhibits pronounced spin-dependent dips in the energy region close to
$\mu$=0.55 eV, resulting from the well-defined spin-dependent Fano antiresonance in the transmission.
On the other hand, narrow spin-dependent Fano dips, which occur in transmission
well below the Fermi energy for Al impurities in the PC configuration do not give visible modifications of the conductance.

When spin mixing takes place on a distance comparable to the systems's length, then only conventional thermoelectric effects can be observed.
In turn, when spin relaxation processes are absent, spin effects become relevant. Below we present numerical results just for this particular case.

Thermal conductance $\kappa_{e}$ of the ferromagnetic nanoribbons with impurities is considerably reduced as compared
to that of the pristine ones. Bearing in mind thermoelectric properties, it can be important that $\kappa_{e}$ is remarkably reduced
in the vicinity of $\mu=0$ for the PM impurity configuration due to the gap occurring for one spin channel. Substantial reduction
of $\kappa_{e}$ is also obtained for negative and positive $\mu$  for one of the PE or PC configurations (Fig. ~\ref{fig7}).
The thermal conductance is additionally reduced for P impurities in the PC configuration due to the well-defined Fano
antiresonance  for chemical potentials close to 0.55 eV.

The influence of impurity atoms on the charge and spin thermopowers is presented in Fig.~\ref{fig8}. For comparison,
the  thermopower
of a pristine nanoribbon is also depicted there. As a general rule one can state that impurity atoms (Al, P) strongly
enhance both charge and spin thermopowers. However, the modifications depend on the position and type of the impurities.
Very remarkable changes can be observed for impurity atoms in the PM configuration, in which -- due to appearance
of energy gap in one of the spin channels -- the charge and spin thermopowers are strongly enhanced in a close vicinity of
$\mu=0$. It should be also noticed that the main contribution to $S_{c}$ and $S_{s}$ in the case of Al atoms comes
from the majority (spin-up) carriers, as this spin channel becomes nonconductive in the presence of Al impurities. On the
other hand, in nanoribbons with P impurities, both $S_{c}$ and $S_{s}$ show different signs, which indicates that the main
contribution corresponds to non-conductive spin-down channel. Totally different results are obtained for pristine
nanoribbons.

Since pristine ferromagnetic nanoribbons exhibit metallic character for both spin directions, the transmission
is constant in the region of small $|\mu|$ and both $S_{c}$ and $S_{s}$ are negligibly small in this region of chemical
potential. Similar behavior can be observed for nanoribbons with impurities in the PE and PC configurations, where both
thermopowers are practically equal to zero for small values of $|\mu|$. However, a considerable enhancement of
S$_{c}$ and S$_{s}$ is then obtained for higher values of $|\mu|$, where pronounced dips occur in transmission
function. High value of $|S|$ in the vicinity of $\mu \approx$ 0.55 eV for  P impurities can be
related to the Fano antiresonance, and thus to rapid changes in transmission function (Fig. ~\ref{fig2}).
As the Fano effect in FM state is strongly spin dependent, the remarkable spin thermopower close to 0.15 mV/K can be
observed in this situation. Strongly enhanced charge and spin thermopowers are also obtained for P impurities localized
near the edges, but they appear for negative $\mu$. The Al atoms in the PE and PC configurations also lead to quite
remarkable modifications. It should be noticed that the edge position of Al atoms leads to considerable enhancement of
$|S|$ for positive $\mu$, whereas impurities localized in the center enhance $|S|$ for negative values of chemical
potential. The opposite relations are found for P  impurities, where  $|S|$ is enlarged for the edge configuration in
the region of negative $\mu$ but for PC configuration the enhancement is for positive $\mu$. All this shows that
type of impurities (Al or P) and their localization in the nanoribbons play an
important role. \newline

\section{Summary and conclusions}

We have considered transport and thermoelectric effects in silicene nanoribbons with Al and P impurity atoms. Using {\it ab-initio} calculations we have determined transmission function through a nanoribbon. Using the calculated transmission we  have determined thermoelectric coefficients in the linear response regime, like Seebeck and spin Seebeck parameters as well as  the electronic contribution to the heat conductance.  The results have been presented for  two different situations corresponding to presence and absence of short-range spin mixing in the nanoribbobs.
The calculations have been performed for both antiparallel (AFM, low energy) and parallel (FM) configurations of the edge magnetic moments.

The key objective was to determine the role of impurity atoms located in various positions with respect to the nanoribbon center. Numerical results clearly show that the Al and P impurity atoms significantly modify the transmission function, and thus also the thermoelectric coefficients. More specifically, the Siebeck and spin Siebeck coefficients are remarkably enhanced by the impurities. This enhancement depends on position of the impurities. We have considered tree different impurity configurations -- PE (impurities at one of the edges), PC (impurities in the center of the nanoribbon), and PM (impurities between the edge and center of the nanoribbon).

\begin{acknowledgments}
This work was supported by the National Science Center in Poland as the Project No. DEC-2012/04/A/ST3/00372.
Numerical calculations were performed at the Interdisciplinary Centre for Mathematical and Computational Modelling (ICM) at Warsaw University and partly at SPINLAB computing facility at Adam Mickiewicz University.
\end{acknowledgments}

\end{document}